\documentclass{acmart}

\usepackage{lscape}
\usepackage{longtable}

\setcopyright{none}
\begin{document}

\title[FAccT-Checked]{FAccT-Checked: A Narrative Review of Authority Reconfigurations and Retention in AI-Mediated Journalism}

\author{Stefano Sorrentino}
\authornote{Both authors contributed equally to the paper}
\email{stefano.sorrentino@epfl.ch}
\orcid{0009-0001-9166-9396}
\affiliation{
  \institution{École Polytechnique Fédérale de Lausanne (EPFL)}
  \city{Lausanne}
  \state{}
  \country{Switzerland}
}

\author{Matilde Barbini}
\authornotemark[1]
\email{matilde.barbini@epfl.ch}
\orcid{0009-0007-7986-2365}
\affiliation{
  \institution{École Polytechnique Fédérale de Lausanne (EPFL)}
  \city{Lausanne}
  \country{Switzerland}
}

\author{Daniel Gatica-Perez}
\email{gatica@idiap.ch}
\orcid{0000-0001-5488-2182}
\affiliation{
  \institution{Idiap Research Institute}
  \city{Martigny}
  \country{Switzerland}
}
\affiliation{
  \institution{École Polytechnique Fédérale de Lausanne (EPFL)}
  \city{Lausanne}
  \country{Switzerland}
}


%% TODO: Replace with your actual author information
%\author{[Your Name]}
%\email{[your.email@institution.edu]}
%\orcid{[your-orcid-id]}
%\affiliation{%
 % \institution{[Your Institution]}
 % \city{[City]}
 % \state{[State/Province]}
 % \country{[Country]}
%}

%% Add additional authors as needed
%\author{[Co-author Name]}
%\email{[coauthor.email@institution.edu]}
%\affiliation{%
%  \institution{[Co-author Institution]}
%  \city{[City]}
%  \country{[Country]}
%}

%\renewcommand{\shortauthors}{[Author et al.]}

\begin{abstract}
The integration of artificial intelligence (AI) into journalistic workflows has intensified concerns about fairness, accountability, and transparency (FAccT). In AI-mediated journalism, these concerns cannot be uniquely understood as properties of technical systems, but instead emerge from deeper reconfigurations of editorial authority, which act as a precondition for FAccT principles. Building on recent interpretivist approaches, we conduct a critical narrative review across journalism studies, human–computer interaction, and FAccT scholarship, conceptualizing editorial authority as the conjunction of decision rights, epistemic warrant, and responsibility. We provide a comprehensive theoretical framework for addressing how concerns on fairness, accountability and transparency emerge, interact, and persist within AI-mediated journalistic practice. 
We identify and describe two concurrent authority reconfigurations driven by AI adoption. First, an internal migration of authority, in which editorial judgment is progressively deferred to large language models (LLMs) embedded within newsroom workflows. This migration occurs not through explicit policy decisions, but through interactional, cognitive, and organizational mechanisms that legitimize AI-generated outputs while obscuring responsibility and weakening individual and professional agency. Second, we analyze an external migration of authority, whereby decision-making power shifts from news organizations toward platforms, vendors, and infrastructural providers that supply AI systems and distribution channels, exacerbating existing power asymmetries within the media ecosystem. Unaddressed, these reconfigurations risk rendering fairness hard to maintain, accountability difficult to assign and transparency performative. 
Additionally, we examine participatory approaches to AI design and deployment in journalism as potential mechanisms for retaining or reclaiming editorial authority. We critically assess both their promise and their structural limitations, highlighting how participation can either meaningfully redistribute authority or function as a tokenistic practice that leaves underlying power relations intact. 
By positioning authority reconfigurations as the central explanatory structure, our contributions reframe these concerns as a problem of institutional power, epistemic legitimacy, and responsibility, offering an holistic foundation for rethinking FAccT interventions in AI-mediated journalism beyond technical optimization alone.

\end{abstract}


%% TODO: Replace with your actual teaser figure if you have one, or remove this section
%  \includegraphics[width=\textwidth]{your-teaser-figure}
%  \caption{Your teaser caption here.}
%  \Description{Description of your teaser figure for accessibility.}

%\received{[Date]}
%\received[revised]{[Date]}
%\received[accepted]{[Date]}

\maketitle

%overview ..we do this employing the narrative review approach -> 1 pagine e mezzo totale 

\section{Introduction}

Information distribution processes have historically been shaped by the early adoption of disruptive technologies \cite{pavlik_impact_2000}. Journalism, in particular, has consistently mediated the relationship between technological innovation and social transformation, whether through the printing press, industrial publishing technologies, or telecommunications, contributing to pivotal historical shifts such as the rise of liberal democracies and the emergence of mass media \cite{sonni_digital_2024, medill2024impact, hassan_journalism_2024}. In the digital age, the trajectory has been no different, and over time algorithmic processes have been fully integrated in journalistic workflows to keep up with the pressure for individualized and commercialized reading experiences \cite{hassan_journalism_2024}, with profound implications for the democratic function of journalism. 

This kind of technological shift affects traditional journalism across multiple dimensions, influencing how journalists perform their work, the characteristics of news content itself, the organizational structure of newsrooms and the broader news industry, as well as the dynamics between news organizations, journalists, and their diverse audiences \cite{pavlik_impact_2000}. These transformations challenge long-standing practices and norms \cite{hassan_journalism_2024, marvin_when_1988}, directly shaping what can be understood as editorial authority: the legitimate power to define, describe, and explain bounded domains of reality through news production and dissemination \cite{tuchman_making_1978}.

Building on journalism's long tradition of technological disruption, the integration of generative AI tools, such as large language models (LLMs), in newsrooms' workflows highlights authority reconfigurations that go beyond adjusting existing routines to new tools, platforms and audiences, calling for a reflection on its core values and implications \cite{hassan_journalism_2024, lewis_generative_2025}. 
Journalists report growing concerns over the reconfiguration of professional authority driven by AI adoption, including the erosion of editorial autonomy, job automation, and the consolidation of technical expertise into specialized knowledge silos within newsrooms \cite{cools_uses_2024, albizu-rivas_artificial_2024, moran_robots_2022, Amigo20042025, tseng_ownership_2025, sonni_digital_2024, xiao_it_2025}. These dynamics shift decision-making power away from journalists toward technical systems and actors, raising uncertainties about fairness, accountability, transparency, and the capacity to exercise editorial judgment, particularly given the opaque nature of foundation models and their tendency to reproduce embedded biases \cite{bommasani_opportunities_2021, bender_dangers_2021, cools_uses_2024}. Such concerns are reinforced by a predominantly top-down, vendor-driven model of AI adoption in newsrooms, with limited journalist involvement in system design and rare in-depth participatory approaches to AI integration  \cite{tseng_ownership_2025, delgado_participatory_2023, suresh_participation_2024}.

In this paper %we make several key contributions to understanding AI's impact on journalism. 
we conduct a critical narrative review \cite{sukhera2022narrative} with the goal of examining how generative AI adoption in digital journalism reconfigures editorial authority as the structural precondition for fairness, accountability and transparency (FAccT) guarantees and meaningful participatory AI implementations. Beginning with the intersection of FAccT concerns in AI-mediated journalism, this narrative approach enables our iterative search process to reveal how these dimensions consistently manifest through struggles over authority, a perspective previously fragmented across disciplinary boundaries.

%we conduct a narrative review to examine how AI adoption, particularly Generative AI, reconfigures editorial authority in digital journalism, addressing three core research goals: (1) understanding how authority operates as a structural precondition for FAccT in journalism; (2) mapping the mechanisms through which authority migrates both internally (to AI systems) and externally (to platforms, vendors, tech companies and roles); and (3) evaluating participatory methods as potential counterforces to authority displacement.

%who holds decision rights, whose knowledge claims receive epistemic warrant, and who bears responsibility for outcomes. 
%This narrative approach enables us to examine FAccT dimensions through an integrated lens of authority, a perspective previously fragmented across disciplinary boundaries.
%We conceptualize editorial authority as the conjunction of \textit{decision rights} (who chooses what gets published), \textit{epistemic warrant} (whose knowledge claims are recognized as legitimate), and \textit{responsibility} (who can be held accountable for outcomes).

Our review yields four contributions:
\begin{itemize}
    \item A \textbf{holistic framework for editorial authority}: We synthesize insights from FAccT research, HCI and journalism studies to provide a comprehensive conceptualization of editorial authority, as the conjunction of \textit{decision rights} (who chooses what gets published), \textit{epistemic warrant} (whose knowledge claims are recognized as legitimate), and \textit{responsibility} (who can be held accountable for outcomes).
    \item A first (to the best of our knowledge) \textbf{in-depth investigation of internal authority migrations in newsrooms under generative AI adoption}: We document how authority progressively relocates from journalists to LLMs through cognitive mechanisms (automation bias, anthropomorphism, normative projection), social dynamics (institutional voice mimicry, perceived alignment), and structural positioning (workflow integration, pseudo-transparency, responsibility gaps).
    \item A \textbf{systematic mapping of external authority migrations in journalism}: We trace how decision-making power relocates from newsrooms to platforms, vendors, and technology companies through infrastructural dependencies, resource asymmetries, and the concentration of AI capabilities, revealing differential vulnerabilities across newsroom scales.
    \item A \textbf{critical evaluation of participatory approaches}: We examine how participatory methods offer potential mechanisms for retaining or reclaiming journalistic authority, while highlighting their conditional nature, implementation challenges, and risks of pseudo-participation when authority reconfigurations are unaddressed.
\end{itemize}

Our findings reveal how these reconfigurations of authority create structural conditions where FAccT concerns become intractable, fairness standards become opaque, accountability diffuses across human-AI assemblages, and transparency obligations lack clear addressees. In the absence of an explicit account of how editorial authority is distributed and exercised in AI-mediated journalism, FAccT analyses risk mislocating these failures at the level of system behavior rather than the institutional power relations that shape them.
%Our main findings reveal that editorial authority in AI-mediated journalism undergoes a dual displacement: internally, through progressive cognitive and structural deference to LLMs that may transform journalists from authors to validators; and externally, through platform dependencies that concentrate decision-making power in technology companies. This reconfiguration creates structural conditions where FAccT concerns become intractable, fairness standards become opaque, accountability diffuses across human-AI assemblages, and transparency obligations lack clear addressees. Participatory methods show promise but remain conditional on early intervention, collective rather than individual forms, and institutional support that is rarely present in current implementations.
To support future research and practice, we also provide: (a) a systematic mapping of practitioners' concerns regarding AI adoption (Appendix \ref{Supplementary Material: Concerns around Generative AI in Digital Journalism}), and (b) a comprehensive atlas of AI functionalities across journalistic workflows (Appendix \ref{Supplementary Material: AI Journalism Atlas - Complete Overview}) , calling for further studies and empirical analyses of authority reconfigurations under AI adoption in diverse journalistic environments.

\section{Methodology}

This paper employs a \textbf{critical narrative review} approach \cite{sukhera2022narrative} to examine authority relocation in AI-mediated journalism. We conducted a narrative review because our investigation required synthesis across distinct literatures, such as journalism studies, HCI, and FAccT scholarship, which systematic review methods would either exclude or force into equivalence. Narrative reviews are appropriate when phenomena are complex and require nuanced interpretation across multiple disciplinary perspectives without predetermined methodological constraints \cite{sukhera2022narrative}, an approach recently adopted also within the FAccT community \cite{10.1145/3715275.3732052}. Although narrative reviews allow interpretive flexibility, we implemented a structured and transparent search and screening procedure to ensure traceability of corpus construction. The introduction of AI into newsroom workflows represents a reconfiguration in journalism's epistemic paradigm, transforming how knowledge claims are produced, validated, and attributed. Traditional systematic reviews, designed to synthesize empirical findings within stable epistemological contexts, cannot adequately capture how the very conditions of knowledge production in journalism are being reconfigured. The narrative review approach, by contrast, enables us to trace authority reconfigurations across multiple levels of analysis, from individual cognitive dynamics to institutional restructuring to platform dependencies.

Editorial authority emerged iteratively as the central analytical construct through sustained engagement with the literature rather than being predetermined. We identified three interconnected patterns of authority reconfigurations: (1) internal migrations, as editorial judgment is deferred to LLMs; (2) external migrations to platforms, vendors, and technology companies through infrastructural dependencies; (3) participatory methods as potential counterforces to retain or reclaim authority. We organized our synthesis around these themes, treating them as the conceptual architecture for interpreting the broader literature, an approach reflecting narrative reviews' interpretive tradition where insights evolve through iterative refinement.

\subsection{Conceptual Framework: Defining Authority as a Necessary Condition to Guarantee FAccT}

Our analysis positions editorial authority in journalism as a structural precondition for FAccT guarantees. This framing emerged from our observation that FAccT concerns in journalism consistently manifested as questions about authority:

\textbf{Authority and Fairness}: Authority determines whose perspectives receive epistemic standing. Assessing fairness requires first identifying who holds authority to define representation standards and whose frameworks structure newsworthiness judgments.

\textbf{Authority and Accountability}: Accountability requires identifiable agents who can be held responsible. Assessing accountability requires first identifying who are the authoritative actors involved.

\textbf{Authority and Transparency}: Transparency obligations follow authority. Determining what transparency is owed, to whom, and for what purpose requires first mapping authority distributions.

Rather than treating fairness, accountability, and transparency as properties to be optimized at the system level, we examine how authority reconfigurations create the conditions under which FAccT concerns might arise.

\subsection{Search Strategy and Study Selection}

Our search proceeded through four phases, searching major scholarly databases, specifically Scopus\footnote{\url{https://www.scopus.com/}}, ACM Digital Library\footnote{\url{https://dl.acm.org/}}, IEEExplore\footnote{\url{https://ieeexplore.ieee.org/}}, Web of Science\footnote{\url{https://www.webofscience.com/wos}}, Jstor\footnote{\url{https://www.jstor.org/}}, Google Scholar\footnote{\url{https://scholar.google.com/}}, as well as grey literature such as newspaper articles, blog posts, technical and technology reports, policy documents, and academic theses. The corpus covers publications from 1955 to 2025, capturing both fundamental literature on sociology of journalism, earlier scholarship on algorithmic and platformized journalism and recent work on generative AI and large language models (see Figure~\ref{fig:distribution} in Appendix \ref{records distribution} for the full distribution by year and thematic category).

Reflecting critical narrative reviews' iterative process, our screening moved from FAccT manifestations in AI-mediated journalism, through the emergence of authority as their underlying structural condition, to editorial authority and editorial judgment as specific analytical objects. 
This process unfolded across four screening phases: (1) identifying work at the intersection of FAccT, journalism, and AI/LLMs; (2) recognizing authority as the underlying analytical condition through backward reference exploration; (3) refining the corpus around editorial authority and journalistic autonomy; and (4) expanding the search to include studies on AI adoption in newsrooms, participatory methods, and platform dependencies shaping editorial power.
Across these phases, studies were retained when they addressed at least one dimension of editorial authority relocation—decision rights, epistemic warrant, or responsibility—in AI/LLM-mediated journalism. The corpus includes empirical, theoretical, critical, and practitioner accounts.

We excluded studies that treated AI solely as a decontextualized technical artifact, including work focused exclusively on model performance, system optimization, or engineering design without engaging the construction of preference data, values, governance assumptions, or their implications for newsroom workflows and accountability. Initial coding identified patterns of authority redistribution that coalesced into the three interconnected themes forming our organizing framework\footnote{The emergent nature of our three themes model reflects the interpretivist foundation of a critical narrative review. Different review teams might identify alternative organizational frameworks. We do not claim our synthesis to be definitive or the only valid interpretation. Rather, we provide a theoretically grounded analysis to support our process model of authority relocation, with explicit acknowledgment that our analytical object and organizing framework emerged through iterative analysis rather than being predetermined.}. Examples of search strings and a detailed description of the screening phases are provided in Appendix~\ref{Supplementary Material: Search Strategy and Screening Phases}. The final corpus yielded 209 records (including one methodology reference \cite{sukhera2022narrative} and 3 records added post-revisions \cite{kellogg2020algorithms, petre2015traffic, Christin2020}).

\section{Related Work: Conceptualizing Authority at the Intersection of FAccT, HCI and Journalism}

%Research on AI in journalism has emerged across several partially overlapping research traditions, including journalism studies, human–computer interaction (HCI), and scholarship within the Fairness, Accountability, and Transparency (FAccT) community. Each of these fields addresses aspects of how algorithmic and AI systems reshape news production, but they often do so using different conceptual vocabularies and analytical priorities. 

By choosing \textbf{editorial authority} as the object of our review, we argue it operates as the structural precondition for fairness, accountability, and transparency guarantees in AI-mediated journalism. This conceptualization emerges from synthesis across three disciplinary traditions that illuminate distinct but interconnected facets of authority's operation and relocation: HCI literature, FAccT research, and journalism studies. 
%This section situates our work within these literatures and highlights how prior research has examined transformations of editorial authority, professional autonomy, and accountability in AI-mediated journalism. 
By bringing these traditions together, our review synthesizes insights that have largely remained distributed across disciplinary venues . %We conceptualize editorial authority as the conjunction of \textit{decision rights} (who chooses what gets published), \textit{epistemic warrant} (whose knowledge claims are recognized as legitimate), and \textit{responsibility} (who can be held accountable for outcomes). This definition emerges from synthesis across three disciplinary traditions that illuminate distinct but interconnected facets of authority's operation and relocation: HCI literature, FAccT research, and journalism studies. 

Existing HCI literature examines authority through interaction design and control allocation, building on foundational frameworks for human-automation interaction \cite{parasuraman2000model} and articulating principles for human-centered AI that emphasize meaningful human control \cite{shneiderman2020human, amershi2019guidelines}. Empirical studies within this tradition operationalize authority through observed patterns of reliance and control in human-AI interaction, documenting users’ overreliance on system recommendations \cite{kapania2022because}, difficulty overriding incorrect outputs \cite{schoeffer2024explanations}, and persistent miscalibration of reliance despite transparency mechanisms \cite{salikutluk2024evaluation, bennett2023does}.

Research on fairness, accountability and transparency reframes authority through responsibility structures and power relations, examining how algorithmic systems assume editorial-like powers while creating accountability gaps through distributed responsibility and opacity \cite{gorwa2020algorithmic}. Algorithmic accountability emphasizes interrogating embedded power relations and investigating impact over outcomes \cite{diakopoulos2015algorithmic}, yet transparency alone proves insufficient when systems resist interpretability \cite{sterz2024quest, knowles_sanction_2021} and social contexts exceed technical disclosure \cite{ananny2018seeing}. Accountability is fundamentally relational, requiring distributed responsibility across development, deployment, and use \cite{santoni2018meaningful}, while avoiding designs and deployments that ignore socio-technical complexity \cite{cooper_accountability_2022, selbst2019fairness}.

Alongside analytical work on algorithmic power, explainability and accountability mechanisms, participatory research traditions in HCI and responsible AI offer a complementary perspective by treating authority and power as outcomes of system design decisions \cite{corbett_power_2023, birhane_power_2022}. Studies on participatory design and participatory AI argue that stakeholders affected by computational systems should be involved in shaping design goals, data practices, and evaluation criteria \cite{delgado_stakeholder_2021}, with different degrees of participation producing different levels of agency and authority \cite{delgado_participatory_2023}. From this perspective, literature on participation offers insights for examining how authority may be retained or redistributed as AI systems are integrated into professional environments.

Authority in journalism, defined as the legitimate power to shape content selection, presentation, and distribution \cite{tuchman_making_1978}, is conceptualized as a contingent, relational achievement. It emerges through interactions between actors claiming authority and audiences recognizing it, grounded in professional legitimacy, identity, norms, and textual practices \cite{carlson_journalistic_2017, carlson_boundaries_2015}. Boundary work, meaning an ongoing discursive process through which definitions of social phenomena are stabilized by drawing boundaries that include certain actors and practices while excluding others, has historically reinforced these claims by distinguishing professional journalism from adjacent practices \cite{carlson_boundaries_2015}; meanwhile, gatekeeping theory systematizes authority across five levels (individual, routines, organizational, institutional, social system), examining how information is selected, ordered, framed, and made visible \cite{shoemaker2009gatekeeping, wallace2017modelling}. In digital environments, professional boundaries, i.e., demarcations of who counts as a legitimate journalist and what practices count as journalism, face intensified contestation as infrastructural and circulation mechanisms play a decisive role in validating knowledge claims \cite{carlson2020journalistic}. Algorithmic systems introduce competing legitimacy bases grounded in "computational objectivity" that devalue human judgment and generate structural tensions \cite{carlson2018automating}, with metrics and analytics infrastructures restructuring newsroom routines and subtly recalibrating professional autonomy under efficiency and market pressures \cite{kellogg2020algorithms, petre2015traffic, Christin2020}.

%Journalists respond through discursive strategies emphasizing irreducibly human qualities (such as trust, empathy and contextual judgment) as foundations for authority that cannot be algorithmically replicated \cite{lewis_et_al_pdf_2019, van_dalen_revisiting_2024, thasler2025journalistic}.

Within this digital journalism context, we conceptualize editorial authority as the conjunction of three interdependent dimensions: decision rights, epistemic warrant and responsibility. %constituting the legitimate power to define, describe, and explain bounded domains of reality through news production and dissemination \cite{tuchman_making_1978}. 
%This tripartite framework bridges journalism studies' concern with professional legitimacy and FAccT scholarship's focus on algorithmic accountability, providing an integrated analytical lens for examining how AI adoption simultaneously reconfigures editorial practice and conditions the possibility of fairness, accountability and transparency guarantees. 
First, \textit{decision rights} refer to institutionally sanctioned power to control information flows through gatekeeping: selecting, ordering, framing, and making visible certain aspects of reality. These rights can be redistributed to algorithmic systems, where editorial curation can happen without explicit recognition \cite{eslami2015always}, shaped by recursive feedback loops in which user behavior both informs and constrains algorithmic outputs \cite{rader2015understanding}. Second, \textit{epistemic warrant} concerns whose knowledge claims are recognized as legitimate, produced through professional credentialing, verification norms, and textual conventions. Epistemic warrant becomes contested when algorithmic systems participate in "truth" determinations, affecting who holds authority to label information as legitimate knowledge \cite{neumann2022justice, hassan_journalism_2024}. Third, \textit{responsibility} concerns who can be held accountable for editorial outcomes, including capacity to interrogate embedded power relations and biases \cite{diakopoulos2015algorithmic}, establish traceability across decision-making \cite{raji2020closing}, and enable contestability by affected parties \cite{vaccaro2021contestability}. 

These research traditions reveal how AI systems reshape journalistic work and institutional power relations, but existing discussions remain fragmented across disciplinary perspectives. Our narrative review synthesizes these contributions through a unified framework of authority reconfigurations and by mapping both the operational deployment of AI across the journalistic workflow (Table \ref{Supplementary Material: AI Journalism Atlas - Complete Overview}) and the professional concerns articulated by journalists regarding AI-mediated journalism (Table \ref{Supplementary Material: Concerns around Generative AI in Digital Journalism}).

\section{Theme 1: Internal Migrations of Editorial Authority}

%Editorial authority, framed as an intersected network of institutional routines, standardized work practices, and individual creative work mediated by the public’s perceptions and demands, defines what constitutes "news" \cite{gans_deciding_2004, tuchman_making_1978, zilberstein_models_2024, carlson_journalistic_2017}. This process operates across interconnected levels, including individual journalists, newsroom routines, organizational structures, institutional frameworks, and broader social systems, through which authority is relationally achieved between actors who claim it and audiences who recognize it \cite{shoemaker2009gatekeeping, carlson_journalistic_2017}. These evolving dynamics rely on performative and strategic rituals, such as the quest for accuracy, reliability, objectivity, and fact-checked truth, aiming at legitimizing institutional news organizations' outputs by embedding them within trustworthy forms and conventions \cite{tuchman1972objectivity, hilligoss_developing_2008, smeenk_journalist_2023, carlson_journalistic_2017, carlson_boundaries_2015}. 
%While these processes have historically been anchored in relatively stable institutional routines and material infrastructures, recent shifts in journalistic practices have reshaped how editorial authority is produced and negotiated internally, generating a set of reconfigurations that are described in the following subsections.
Editorial authority emerges from an intersected network of institutional routines, standardized practices, and individual creative work, mediated by public perceptions and demands that define what counts as "news" \cite{gans_deciding_2004, tuchman_making_1978, zilberstein_models_2024, carlson_journalistic_2017}. %Operating across interconnected levels, including individual journalists, newsroom routines, organizational structures, institutional frameworks, and broader social systems, authority is relationally achieved between actors who claim it and audiences who recognize it \cite{shoemaker2009gatekeeping, carlson_journalistic_2017}. 
These dynamics rely on performative and strategic rituals such as accuracy, objectivity, and fact-checking, aimed at legitimizing institutional news organizations' outputs by embedding them within trustworthy forms and conventions \cite{tuchman1972objectivity, hilligoss_developing_2008, smeenk_journalist_2023, carlson_journalistic_2017, carlson_boundaries_2015}.
While historically anchored in stable routines and infrastructures, recent shifts in journalistic practice have reconfigured how editorial authority is produced and negotiated internally, as discussed in the following subsections.

\subsection{Digital Transformation and the Emergence of Socio-Technical Authority } \label{digital transformation}

Traditionally, editorial authority has been mediated through relatively closed, human-centered assemblages, where journalists operated within routines and hierarchies as the primary agents responsible for translating institutional decisions into published news. 
%Traditionally, journalistic routines have been mediated between individual human agents, including reporters, editors and sources, whose interpretive labor gave concrete form to institutional standards and professional norms \cite{smeenk_journalist_2023, carlson_journalistic_2017}. Through practices of sourcing, verification, framing, and narrative construction, journalists operated as the primary agents responsible for translating institutional authority into published news. %The material infrastructure of print journalism further played a central role in stabilizing this configuration of authority: printed newspapers fixed editorial choices in time and form, rendering news as a durable, bounded artifact that appeared as settled, complete and reliable information \cite{hayles_2002_print}. Print anchored institutional voice to identifiable human agents and editorial hierarchies, reinforcing the perception of news as a finalized and authoritative account of reality. In this sense, writing functioned as the medium of "truth", granting permanence and legitimacy to the institutional judgments embedded in journalistic text, reinforcing the authority definition process \cite{hassan_journalism_2024}.
The material infrastructure of print journalism contributed to stabilizing this configuration by fixing editorial choices in time and form, anchoring institutional voice to identifiable actors and hierarchies, lending durability and legitimacy to journalistic texts \cite{hayles_2002_print, hassan_journalism_2024}. 
The shift from print to digital and the progressive integration of algorithmic tools in editorial workflows brought non-journalistic actors into the news-making process \cite{carlson_boundaries_2015, zilberstein_models_2024}, profoundly disrupting such routines and reconfiguring authority. From a media-theoretical perspective, authority and agency became more explicitly distributed across socio-technical assemblages of humans, institutions, and machines \cite{deuze2008changing, hayles_cognitive_2016, hayles_2002_print, hayles_how_2010, bomba_agency_2025, elish_moral_2019, tsamados_human_2025}, with sense-making increasingly mediated through digital interfaces and infrastructural forms of authorship \cite{punday_2025_digital}. As news moved to networked digital platforms, information became disembodied from stable material forms, circulating as mutable content \cite{hayles_2002_print, hayles_how_2010}. This move weakened the link between authorship, accountability, and institutional voice, rendering journalistic texts less clearly bounded and autonomous subjects less defined, forming heterogeneous informational entities with continuously redefined boundaries \cite{bomba_agency_2025, carlson_boundaries_2015}. %As news moved from print to networked digital platforms, information became increasingly disembodied from stable material forms, circulating as mutable and updateable content \cite{hayles_2002_print, hayles_how_2010}. This loss of material fixity weakened the link between authorship, accountability, and institutional voice, rendering journalistic texts less clearly bounded in time, form, and authorship. As a result, autonomous subjects and authors participating in the authority definition process became less defined, leading to a collection of heterogeneous components forming an informational entity whose boundaries undergo continuous re-definition \cite{bomba_agency_2025, carlson_boundaries_2015}. 

These changes have reconfigured modern newsrooms as socio-technical collectives, in which news production now emerges through a joint effort between humans, institutions and computational systems \cite{seaver_algorithms_2017, carlson_robotic_2015, diakopoulos_automating_2019, thasler2025journalistic}. Digital tools have increasingly exerted influence within editorial processes by being legitimized through their integration into pre-established work routines, mediating human agency and shaping the conditions under which journalistic judgment is exercised \cite{dodds_controlled_2025}. Recommendation engines, content management systems, and automated reporting tools exemplify this shift towards algorithmic power, structuring visibility, value, and publishability %according to audience metrics and market logics 
\cite{diakopoulos_automating_2019, diakopoulos2015algorithmic}, with metrics and analytics tools reorganizing newsroom priorities and professional norms, embedding market-oriented logics into everyday editorial decision-making \cite{kellogg2020algorithms, petre2015traffic, Christin2020}. %Automated recommendation engines, content management systems, and automated news reporting tools exemplify this shift, structuring what is visible, valuable and publishable according to audience metrics and market logics through practices of prioritization, classification, association, and filtering that constitute a distinctive form of algorithmic power \cite{diakopoulos_automating_2019, diakopoulos2015algorithmic}.

While acquiring epistemic relevance within journalistic workflows, such tools remained largely procedural, conditioning editorial judgment rather than directly producing it, operating in a supporting capacity within pre-established routines and reflecting an editorial logic grounded in subjective judgments by authorized professional experts.
In contrast, the integration of LLMs into editorial workflows marks a qualitative shift in this configuration that defines institutional authority. Unlike prior systems, LLMs participate directly in the formation of journalistic outputs, acting as proxies for editorial judgment and becoming epistemic actors within institutional routines \cite{tsamados_human_2025, bomba_agency_2025, shin_automating_2025} contributing to the "web of facticity" \cite{tuchman_making_1978} that frames news production. Such uses include brainstorming companions, assisted news-angle generation, source monitoring, headline suggestions, automated writing and creative supports\footnote{see Table in Appendix \ref{Supplementary Material: AI Journalism Atlas - Complete Overview} for an overview of current generative AI uses across journalistic practices} \cite{dodds_controlled_2025}. Generating institutionally legible narratives and interpretive outputs, these systems intervene more directly in the epistemic processes through which journalistic authority is constituted \cite{diakopoulos_automating_2019, thasler2025journalistic}. 
We argue that this reconfiguration calls for new reflection on how fairness, accountability, and transparency are addressed when authority is no longer mediated solely through human agency in editorial judgment.

\subsection{Perceived Authority and Social Legitimacy of AI in Newsrooms } \label{perceived authority}

Research from HCI is relevant to understand how generative AI tools further acquire perceived epistemic status \cite{shin_automating_2025} exhibiting interpretative traits that align with established professional, journalistic and social habits, such as fluency, confidence, and role performance \cite{smeenk_journalist_2023, zilberstein_models_2024}. 

In hybrid socio-technical collectives like modern digital newsrooms, the CASA (Computer as Social Actors) paradigm \cite{nass_computers_1994} describes how computational systems become legitimized actors, not only as a result of their integration in professional routines and institutional norms, but because they manifest social cues. LLMs produce even more explicit and stronger social cues, exhibiting stylistic competence and politeness, to which humans unconsciously respond applying social rules and expectations, even when aware they are interacting with machines. The "media equation" \cite{reeves1996media} provides further insights about these automatic cognitive mechanisms through which humans perceive entities that project social as potential collaborators, other than tools. As a result, LLMs function as "anthropomorphic agents" where human-like interaction constitutes an intrinsic feature \cite{peter2025benefits, norman_how_1994}, generating outputs 
that simulate the attributes through which editorial authority is shaped, recognized and valued: adjustment, intentionality, and judgment \cite{hilligoss_developing_2008, norman_how_1994}. 

Deference behaviors are predisposed by unconscious automation biases in users’ cognitive processes, leading them to defer to system outputs, assuming that computational processes are more objective or reliable than human judgment \cite{liel2025turning, parasuraman_complacency_2010, shin_automating_2025, lai_towards_2021, sundar2019machine, parasuraman_humans_1997}. 
In professional contexts, automation bias intensifies when LLMs produce text that appears confident, coherent, institutionally disciplined, and aligned with norms and standards \cite{khamassi_strong_2024}, corresponding to journalists' concerns about the erosion of editorial judgment, accountability, accuracy and trust\footnote{see Table in Appendix \ref{Supplementary Material: Concerns around Generative AI in Digital Journalism} for an overview of professional concerns about generative AI in journalism}. AI-generated news content presents the structural hallmarks of legitimate journalism \cite{lermann2024effects}, mimicking surface-level markers of journalistic credibility. Lexical, syntactic, and structural conventions associated with institutional voice are reproduced, including editorial vocabulary (e.g., "according to sources"), passive constructions that create objectivity effects \cite{smeenk_journalist_2023}, and information hierarchies reflecting entrenched news values. Most consequentially, LLMs can conform to the specific linguistic conventions of particular news organizations through training data or instruction prompting \cite{punday_2025_digital}. This institutional voice mimicry encourages normative projection, whereby users infer that outputs embody the professional values they recognize \cite{akbulut2024all, seaver_algorithms_2017}. Such projection is not irrational in newsroom contexts, where stylistic conformity typically signals editorial review and normative compliance \cite{hilligoss_developing_2008}. The result is an institutional voice that appears normatively aligned with the performative rituals that make journalism recognizable \cite{zhou2023synthetic, smeenk_journalist_2023, zilberstein_models_2024}.
%Lexically, LLMs reproduce editorial vocabulary: "according to sources," "officials confirmed," "data shows." Syntactically, they employ the passive constructions and attributive phrases that create rhetorical distance and objectivity effects \cite{smeenk_journalist_2023}. Structurally, they organize information according to news values hierarchies embedded in their training data. Most consequentially, they generate outputs that conform to the specific linguistic conventions of particular news organizations (either because those conventions pervade the training corpus or through instruction prompting) \cite{punday_digital_nodate}. The result is an institutional voice that appears normatively aligned with the performative practices and rituals that make journalism recognizable \cite{zhou2023synthetic, smeenk_journalist_2023, zilberstein_models_2024}. 

This "implicit alignment" perception is amplified by traditional training techniques. Technical alignment in AI, referred to the degree to which a machines' optimization matches intended human values and behaviors, is increasingly labeled as "weak", especially with approaches such as reinforcement learning from human feedback (RLHF) \cite{khamassi_strong_2024, gabriel_artificial_2020}. Although RLHF and similar methods fine-tune models to produce outputs deemed helpful, harmless, and honest by annotators (often operating outside specific newsrooms), this optimization emphasizes linguistic plausibility over genuine internalized values \cite{mazeika_utility_2025}, further advancing the performative capacities of models. As a result, models can convincingly emulate institutional voices and routines without truly embodying their values, creating a significant asymmetry between the "strong" perceived institutional alignment and the "weak" technical alignment \cite{komatsu2020ai}. This perception gap, emerging from institutional voice mimicry \cite{seaver_algorithms_2017}, is further amplified by cognitive heuristics. The machine heuristic \cite{sundar2019machine} leads users to treat machine-generated outputs as inherently objective, attributing neutrality to computational provenance \cite{hayles_how_2010, hilligoss_developing_2008}. In journalism, this heuristic is particularly potent, as objectivity functions as a core professional ideal that algorithmic systems appear to embody \cite{jakesch2023co}. In fast-paced newsroom environments, efficiency pressures, deadlines, and cognitive overload further encourage reliance on such heuristics, increasing the likelihood that journalists interpret LLM outputs as expressions of journalistic expertise rather than statistical pattern matching \cite{liel2025turning, parasuraman_complacency_2010, shin_automating_2025}, as automation adoption under conditions of resource scarcity and productivity pressure can gradually recalibrate professional judgment toward efficiency and measurability \cite{kellogg2020algorithms, petre2015traffic, Christin2020}. 

%Prior studies show how automation adoption in journalism under conditions of resource scarcity and productivity pressure gradually recalibrates professional judgment toward efficiency and measurability, making algorithmic guidance increasingly difficult to resist (Christin AND Petre). 

This performative legitimacy complicates existing frameworks of accountability, as editorial authority is inferred from stylistic alignment rather than traceable editorial processes. For HCI and journalism studies, this reconfiguration calls for renewed attention to how responsibility and transparency can be articulated when epistemic authority is mediated by systems that simulate professional judgment.

\subsection{Responsibility Gaps and the Reconfiguration of Journalistic Agency } \label{responsibility}

Interactional trust and implicit alignment are sustained and amplified by the opacity and limited legibility of LLMs' decision-making processes \cite{seaver_algorithms_2017, hilligoss_developing_2008}, combined with pseudo-transparency mechanisms that create false impressions of accountability. When an LLM generates a particular framing for a story, selects certain sources over others, or emphasizes specific angles, the "reasoning" behind these choices remains largely inaccessible to human review. This opacity creates "responsibility gaps" \cite{matthias2004resp}: situations where harmful outcomes cannot be attributed to any specific agent because causation is distributed across multiple actors and systems \cite{bomba_agency_2025, elish_moral_2019, tsamados_human_2025}. In newsroom contexts, when an LLM-generated article contains subtle bias, privileges certain perspectives, or omits critical context, determining responsibility becomes intractable. 

Attempts to address opacity through Explainable AI (XAI) mechanisms have often  proven insufficient or counterproductive, creating pseudo-transparency that reinforces (rather than mitigates) the perceived implicit alignment. Feature-based explanations can exacerbate over-reliance by creating false confidence in understanding, altering reliance patterns independent of correctness, instead of helping humans discern correct versus incorrect AI recommendations \cite{schoeffer2025ai, romeo2025exploring}. In practice, user judgment of AI outputs is strongly influenced by how the system and its recommendations are described, which can create misleading confidence in the outputs \cite{chu_user_2026, luger_like_2016}. In newsroom contexts, this effect becomes consequential when explanations present normative rationales for editorial choices, such as claims that a framing "serves the public interest" or that selected sources "ensure balance." Rather than documenting decision processes, such explanations function as justificatory narratives that absorb uncertainty and stabilize trust. Pseudo-transparency thus produces a "moral buffer" \cite{carnat2024human}, in which users experience psychological distance from outputs they did not author but feel they have sufficiently understood and validated \cite{elish_moral_2019, khamassi_strong_2024}. %This is particularly risky in journalism, where LLM providing explanations like "this framing emphasizes public interest" or "these sources provide balanced perspective" perform and mimic institutional alignment rather than documenting it. Pseudo-transparency through XAI thus creates a "moral buffer" \cite{carnat2024human}: users experience psychological distance from outputs they did not fully author but feel they understand and have validated \cite{elish_moral_2019, khamassi_strong_2024}. 
%This buffer can facilitate two forms of responsibility avoidance. First, human operators defer: journalists may defend problematic outputs by noting that they followed the LLM's recommendation, which appeared reasonable and reliable given its explanation \cite{fritz_deference_2025}. The system's apparent transparency creates a basis for delegating blame. Second, institutions shift accountability: news organizations point to human-in-the-loop protocols, arguing that journalists retain final authority and bear responsibility for any failures \cite{dodds_controlled_2025}. Such institutional framing obscures how workflow design, cognitive biases, and pseudo-transparency combine to undermine meaningful oversight \cite{elish_moral_2019, liel2025turning, norman_how_1994, tsamados_human_2025}. 
This buffer facilitates responsibility avoidance in two ways. First, journalists may defer responsibility by appealing to apparently reasonable system recommendations, citing explanatory rationales as evidence of diligence \cite{fritz_deference_2025}. Second, institutions may shift accountability by invoking human-in-the-loop protocols that nominally preserve editorial control while obscuring how workflow design, cognitive bias, and pseudo-transparency undermine meaningful oversight \cite{dodds_controlled_2025, elish_moral_2019, liel2025turning, norman_how_1994, tsamados_human_2025}.

The convergence of these mechanisms ultimately produces an \textbf{internal migration} to the model: \textit{the progressive transfer of editorial authority from journalists to LLMs through cognitive, social, and structural channels that operate largely outside any deliberate institutional awareness.} This migration is "internal" as it occurs within the newsroom workflow, without explicit external imposition. Unlike platform algorithms that determine distribution or advertising systems that influence content through revenue pressure, LLMs are integrated directly into editorial production as assistants, with their authority emerging from within the writing process itself. Consequentially, the migration is "internal" in the sense of being psychological and cultural rather than formal \cite{seaver_algorithms_2017}, as no newsroom policy explicitly grants LLMs editorial authority, yet that authority accumulates through the mechanisms we have analyzed as core components of modern hybrid socio-technical collectives: (1) automation bias predisposes deference \cite{lai_towards_2021}; (2) anthropomorphism and CASA dynamics confer quasi-colleague status \cite{peter2025benefits}; (3) institutional voice mimicry through "weak" technical alignment creates a "strong" perceived professional alignment; (4) normative projection fills gaps in actual values \cite{jakesch2023co}; (5) workflow positioning structures cognitive labor to increase reliance in AI outputs \cite{romeo2025exploring}; and (6) opacity and pseudo-transparency dissolve accountability \cite{carnat2024human}. Each mechanism operates incrementally, with their combination producing a qualitative transformation, as LLMs transition from tools to systems that exercise editorial authority within a profoundly hybrid professional environment \cite{hayles_how_2010, shin_automating_2025, hayles_cognitive_2016}.

Empirical indicators of this migration are already visible. Studies have documented journalists who conform their work to algorithmic preferences despite knowing that systems are unreliable, in a form of "algorithmic conformity" where normative pressure to use the tool for "efficiency" overrides professional judgment \cite{liel2025turning, fischer2025efficiency}. Furthermore, the finding that an increasing percentage of newsrooms deploy AI for writing, analysis, and personalization \cite{tseng_ownership_2025} suggests that internal migration is no longer exceptional, but typical. More broadly, HCI research indicates that domain experts, including journalists, provided with pre-populated algorithmic suggestions generate fewer novel insights, implying how algorithmic output can constrain, rather than expand, human cognitive contribution \cite{spangher-etal-2024-llms, levy2021assessing, parasuraman_complacency_2010}, as rejecting or substantially revising AI-generated content often requires more cognitive effort than producing it from scratch, creating path dependencies in which the model’s initial framing shapes the final output \cite{thasler2025journalistic}.
In journalistic environments, a possible completion of internal migration hence occurs when LLMs effectively function as editorial authorities whose judgments implicitly frame subsequent human deliberation rather than merely assisting it.

From an authorship and agency perspective, this migration calls for a reflection on the reconfiguration of what authorship itself means for concerned journalists and accomplishes in AI-ready newsrooms\footnote{see Table in Appendix \ref{Supplementary Material: Concerns around Generative AI in Digital Journalism} for an overview of professional concerns about generative AI in journalism}. When editorial authority migrates towards LLMs, the journalist's role transforms from author to validator, from sense-maker to quality-checker and selector \cite{bomba_agency_2025, dodds_controlled_2025, smeenk_journalist_2023, zilberstein_models_2024, thasler2025journalistic}. 
%The cognitive effort required to reject or substantially revise AI-generated content is asymmetrically higher than creating it from scratch, producing path dependencies where initial AI framing determines final output. Authority migrates to the model not through demonstrated competence but through strategic positioning within institutional routines that implicitly delegate judgment \cite{thasler2025journalistic}. 
As a function that stabilizes accountability, responsibility, authority, and legitimacy within institutional regimes of knowledge \cite{foucault1977what, foucault1980power, foucault_truth_2009}, authorship persists symbolically while being weakened operationally, a condition that echoes earlier critiques of the modern author as a figure that no longer centers meaning, intention, and responsibility within a singular subject \cite{barthes_death_1997}. This ultimately leads to an increase of epistemic judgments shaping news narratives that resist attribution and contestation, beyond individual journalists as named authors and formal decision-makers \cite{bomba_agency_2025, dodds_controlled_2025, hayles_how_2010, zilberstein_models_2024}. 
In summary, editorial authority is reconfigured through an internal migration: journalists retain accountability for published content while exercising diminished control over the processes and tools that generate it. This weakening of the author function marks a shift both in how journalistic authority is produced and in what it can meaningfully sustain.

\section{Theme 2: External Migrations of Editorial Authority}

\label{sec:external_migration}
%Our analysis of the literature reveals that the internal reconfigurations of authority described in the previous section are complemented by equally relevant external migrations. While scholarship has examined various aspects of how platforms, vendors, and technology companies exert influence over journalism, including platform dependencies \cite{nielsen2022power, chua2022platform}, algorithmic governance \cite{gillespie2014relevance, papaevangelou_funding_2024}, and infrastructure provision \cite{papaevangelou_funding_2024, simon_uneasy_2022}, these insights have frequently remained fragmented across disciplinary boundaries \cite{ausserhofer_et_al_datafication_2020, anderson2013towards, coddington_clarifying_2015, oh2025harmonizing, dodds2024impact}. We provide a comprehensive synthesis of external authority migration that, examined alongside internal migration, reveals how AI adoption simultaneously displaces authority upward to platforms, vendors, and technologists and inward to algorithmic systems, creating compound dependencies that fundamentally reconfigure the conditions under which journalism operates.
Our literature analysis reveals that internal reconfigurations of authority are mirrored by external migrations. Research has examined how platforms, vendors, and technology companies shape journalism through platform dependencies \cite{nielsen2022power, chua2022platform}, algorithmic governance \cite{gillespie2014relevance, papaevangelou_funding_2024}, and infrastructure provision \cite{papaevangelou_funding_2024, simon_uneasy_2022}, but these insights often remain scattered across disciplines \cite{ausserhofer_et_al_datafication_2020, anderson2013towards, coddington_clarifying_2015, oh2025harmonizing, dodds2024impact}. Our synthesis of \textbf{external authority migration}, considered alongside internal shifts, shows how \textit{AI adoption simultaneously displaces authority upward to platforms, vendors, and technologists and inward to algorithmic systems, producing compound dependencies that reshape journalism’s operating conditions.}

\subsection{Power Structures and Asymmetries in News Media Environments}
\label{subsec:foundation_power}

Newsrooms have long been organized around hierarchical power relations, with editors and publishers setting policies and controlling resources, while reporters exercise uneven and status-dependent autonomy; as a result, subtle forms of organizational power and social control shape both news selection and the broader news product \cite{pavlik_impact_2000,breed_social_1955}. Although editorial independence is often framed as a cornerstone of press freedom, in practice it is conditional and contested: news organizations remain dependent on funding, access to information, infrastructures, power relations, and technology; consequently, debates over independence often hinge on whether political, financial, or technological pressures are seen as most problematic \cite{paik_journalism_2025,hamada2019editorial,jane_b_singer_contested_2007,karppinen_what_2016,humprecht_ownership_2019,drunen_safeguarding_2023,cushion_data_2017,stalph_classifying_2018,ananny_2018_freedom,schudson_virtues_2005,simon_uneasy_2022,simon_escape_2024,simon_artificial_2024,cools_uses_2024}.

Technological change has reshaped editorial control by redistributing authority beyond the newsroom. Automated systems embed external values into journalistic decisions, transferring influence to data providers, personalization tools, and algorithm designers \cite{drunen_safeguarding_2023,gillespie2014relevance,jones_ai_2022,becker_policies_2025,farrell_moral_2023,stalph_classifying_2018,malcorps_news_2019,kunert_form_2019,bodo_selling_2019,Kitchin02012017,dierickx_pdf_2021,outi_lundahl_algorithmic_2022}. 
Historically, personalization systems and audience analytics introduced quantitative logics into newsrooms, operationalizing metrics as editorial criteria \cite{Christin2020, petre2015traffic} and creating new \emph{power loci} \cite{beckett2019new,coddington_clarifying_2015,stalph_classifying_2018,tabary2016data,malcorps_news_2019,diakopoulos2014algorithmic,turow2013daily,drunen_safeguarding_2023} where gatekeeping authority could shift from journalists to the technologists and commercial teams who design these systems \cite{kellogg2020algorithms, petre2015traffic, Christin2020}. At the same time, digital communication has fostered direct audience interaction that influences story selection and coverage \cite{pavlik_impact_2000}. These shifts have brought new professional actors into journalism, with algorithms and humans working in complementary roles. IT specialists, data analysts, and algorithm developers increasingly occupy hybrid roles as editorial technologists \cite{lischka2023editorial, oh2025harmonizing}, integrating computational expertise into news production. However, their growing involvement has also generated tensions as they challenge established authority and resource control \cite{diakopoulos2020computational,anderson2013towards,moran_robots_2022,kosterich2020managing,lewis2015actors,thurman_algorithms_2019}.

External authority migrations do not affect all newsrooms equally. The distribution of power in news media organizations varies significantly by scale, creating differential vulnerabilities to external dependencies. Large organizations manage complexity through hierarchies and delegation of decision-making processes~\cite{skrastins_how_2019}, and such hierarchies tend to persist even in contexts designed for more balanced power relations, as reductions in formal structures are often offset by the rise of informal ones~\cite{schuster_power_2024}. Hierarchies can indeed serve functional purposes in large settings, supporting motivation, coordination, and conflict reduction~\cite{schuster_power_2024}, yet excessive reliance on centralized authority may hinder exploration, learning, and resilience, for example when vetoes from less knowledgeable layers suppress beneficial change~\cite{dosi_hierarchies_2021}. By contrast, small organizations are especially vulnerable to power asymmetries, as larger stakeholders (e.g., digital platforms and corporate conglomerates such as large retail chains or buyer partners) can impose contractual controls \cite{drunen_safeguarding_2023}, exploit high technological costs \cite{simon_escape_2024}, and lock smaller players into dependencies that undermine their autonomy and values \cite{oldham_navigating_2025, jones2023generative}. Research shows that organizations with greater financial, technical, and human resources are better positioned to maintain independence, while smaller or local ones often struggle to compete, adapt, or resist external pressures~\cite{oldham_navigating_2025,jones2023generative,simon_escape_2024,drunen_safeguarding_2023}. These resource gaps are further exacerbated by structural dynamics such as ownership concentration, which increases inequalities in the media sector, making it especially difficult for smaller outlets to sustain high-quality journalism~\cite{humprecht_ownership_2019}. This differential capacity becomes critical when understanding how AI adoption accelerates the external migration of authority, as smaller newsrooms lack the resources to resist platform dependencies that larger organizations can sometimes negotiate or circumvent.

\subsection{The Mechanisms of External Migrations: Platform Dependencies and Infrastructural Capture}
\label{subsec:platform_mechanisms}

 Migrations of authority from newsrooms to external actors operate through several interconnected mechanisms, with platform dependencies serving as the primary conduit. By positioning themselves as providers of AI tools and infrastructure (see Table \ref{Supplementary Material: AI Journalism Atlas - Complete Overview} for comprehensive mapping of AI functionalities), platforms extend their role from being primarily channels of connection to also serving as production infrastructures. Such reliance complicates accountability, as publishers find it difficult to challenge or regulate the very companies they depend on for audience reach, distribution, and monetization~\cite{bell_platform_2017}, creating contingent negotiations in which platforms set most of the terms despite some room for counterbalancing strategies~\cite{nielsen2022power,poell2023spaces,chua2022platform}. The resulting asymmetry is relational but persistent: publishers remain cautious of over-dependence, yet their content and legitimacy also sustain platform ecosystems. Funding mechanisms and infrastructural advantages deepen this imbalance~\cite{papaevangelou_funding_2024}, with high fixed costs of frontier AI concentrating effective control in a small number of organizations. While platform dependence is not new for news organizations, AI adoption intensifies these asymmetries by adding production infrastructure dependencies to existing distribution and monetization ones. This newsroom-level dependency reflects a broader structural dynamic that some scholars describe as "algocracy" \cite{aneesh2006virtual}, capturing the structural risk of embedding authority in technological systems, which centralizes decision-making power in technical elites and sidelines under-represented populations, including the local communities that smaller outlets aim to serve~\cite{corbett_power_2023}. In practice, a small group of companies provides the infrastructures and datasets on which much of the news media sector depends, producing highly uneven access to advanced AI capabilities across different regions of the world~\cite{hansen_et_al_initial_2023}.

%These asymmetries manifest in distinct ways within news organizations. Within corporate chains, technological systems are deployed to maximize throughput, consolidate roles, and cut skilled labor, often reducing the time available for verification and weakening the quality of local watchdog reporting~\cite{higgins-dobney_news_2021}. Smaller newsroom editors, facing financial and staffing pressures, describe AI adoption as a survival strategy, even when it entails ceding elements of editorial control or compromising professional standards. Platform partnerships and AI-driven distribution further reinforce this dependency, raising ethical dilemmas about who ultimately determines journalistic norms~\cite{paik_journalism_2025}. Furthermore, when under-resourced newsrooms adopt external AI to process local data, they encounter tensions around privacy, data governance, transparency, and accountability: in these contexts, limited capacity often means trading away elements of trustworthy AI in order to gain access to basic functionality~\cite{quere_trust_2022}.

At the systemic level, these dependencies generate feedback loops that further entrench platform power: each use of platform tools generates data that enhances the platforms' own models, allowing them to assume functions once central to journalism and reinforcing existing asymmetries~\cite{simon_escape_2024,jungherr2021digital}. Editorial judgment and transparency narrow as opaque systems mediate production and distribution, creating an "intelligibility problem" that threatens journalistic values and professional quality~\cite{simon_escape_2024,jones_ai_2022}, concerns practitioners consistently identify regarding transparency, accountability, and editorial autonomy (Table~ \ref{Supplementary Material: Concerns around Generative AI in Digital Journalism}). External monetization is an additional concern: the very act of using platform AI strengthens the competitive position of technology companies, enabling them to undercut publishers' business models and shifting power further toward infrastructural incumbents~\cite{simon_artificial_2024}.

\subsection{Newsrooms' Reconfigurations as Consequence of External Dependencies}
\label{subsec:internal_reconfigurations}

External migrations of authority produce reconfigurations within newsrooms themselves and, while these changes occur internally to the organization, we argue they represent an externalization of authority from journalists toward technical systems and specialized roles. The integration of LLMs, and AI in general, fundamentally redistributes editorial authority, shifting power from traditional editorial roles toward technological systems and new technical positions. This redistribution occurs as responsibilities once held by editors are partially transferred to managers and technical staff, while journalists become increasingly bound to digitized workflows that narrow their autonomy~\cite{garcia_aviles_journalists_2004,dodds2024impact}. These migrations are reinforced by the opacity of AI systems and the separation between engineering and editorial units, which foster knowledge silos and reliance on technical specialists ~\cite{cools_news_2024,dodds2024impact}. As a result, journalists often defer AI responsibility to technical or legal teams, and this fragmentation of accountability, combined with weak cross-departmental coordination, can entrench bias when non-technical staff cannot contest system behavior~\cite{dodds2024impact,dorr2017ethical}, contributing to the concerns documented in Table \ref{Supplementary Material: Concerns around Generative AI in Digital Journalism}. 
In response to these external reconfigurations of authority, journalists defend professional autonomy by framing uniquely human qualities such as judgment, empathy, and creativity as central to their work~\cite{carlson_robotic_2015,van_dalen_revisiting_2024,schapals_assistance_2020}. At the same time, their judgment and autonomy is increasingly constrained as audience metrics quantify editorial choices and algorithmic optimization steers both content production and its distribution~\cite{bodo_selling_2019,paik_journalism_2025}. As a result, editorial independence, long central to journalistic identity, now extends beyond resisting political and commercial pressures to also encompass the ability to shape how automated systems embed and enact editorial values~\cite{hamada_determinants_2022,reich_and_hanitzsch_full_2013,van_drunen_editorial_2021}.

%\label{subsec:journalist_perspectives}

\section{Theme 3: Editorial Authority Retention Through Participatory Methods in AI-Mediated Journalism} \label{Participatory methods and AI}

As editorial authority in AI-mediated newsrooms is progressively displaced through internal cognitive delegation and external infrastructural dependence, journalists articulate concerns about the erosion of editorial autonomy, intellectual property, data protection and the concentration of technical expertise that marginalize editorial judgment\footnote{see Table in Appendix \ref{Supplementary Material: Concerns around Generative AI in Digital Journalism} for an overview of professional concerns about generative AI in journalism}, explicitly calling for more meaningful forms of involvement in the AI integration process  \cite{cools_uses_2024, albizu-rivas_artificial_2024, moran_robots_2022, Amigo20042025, tseng_ownership_2025, sonni_digital_2024, xiao_it_2025}.
FAccT and HCI research has focused particularly on participatory methods as ways to intervene in those mechanisms that contribute to authority reconfigurations by counteracting opaque workflow integrations, disrupting biases, narrowing the responsibility gaps, and empowering the function of newsrooms and journalists as  actors in the AI adoption process. Drawing on this tradition, participatory design literature provides insights on how authority can be retained in ways that preserve decision rights, epistemic warrant, and responsibility under human-AI hybrid conditions \cite{cooper_systematic_2022, delgado_participatory_2023, suresh_participation_2024, bannon_reimagining_2018}. 

\subsection{Degrees of Participation and Their Impact on Editorial Authority } \label{degrees of participation}

Research consistently shows that participation is not a binary condition, but a graded structure where different forms of involvement produce various degrees of agency, ownership, and authority \cite{arnstein_ladder_1969, delgado_participatory_2023, suresh_participation_2024, delgado_stakeholder_2021, ajmani_secondary_2025}. Minimal, performative participation rarely alters power relations, whereas deeper forms of contribution that intervene upstream in system design, data selection, and evaluation criteria can meaningfully redistribute decision rights and epistemic control \cite{corbett_power_2023, birhane_power_2022, palacin_design_2020, cooke_participation_2001, sloane_participation_2022}. In the context of journalism, this distinction plays an important role: participation that only allows journalists to approve or edit AI model outputs leaves the core authority structure intact, reproducing the validator role described in section \ref{responsibility} rather than counteracting it, while more in-depth participation roles can effectively intervene in the internal migrations of authority \cite{delgado_participatory_2023, simon_uneasy_2022, xiao_it_2025, thasler2025journalistic}.

To achieve meaningful degrees of participation that move beyond passive roles for journalists, collaborative AI projects in newsrooms should address what Suresh et al. define as "participatory ceiling" \cite{suresh_participation_2024}, i.e., the inherent limit of a community to impact external "one-size-fits-all" models. In this context, Tseng el at. propose the Newsroom Tooling Alliance (NTA) framework as a way to operate at a "subfloor" level (as part of a blueprint in which the "foundation" level is represented by the adopted commercial AI model and the "surface" level is the context-specific model optimization) to address macro (market dynamics) and meso (inter-organizational) tensions \cite{tseng_ownership_2025}, described in section \ref{subsec:platform_mechanisms}. This structural participation allows journalists to preemptively and collectively set the terms of engagement, reclaiming control over data usage and functionality before these systems are integrated into fixed newsroom routines, through a co-designing approach where decisions are negotiated in a shared space. 

%automation bias:

We further argue that different degrees of participation also shape how authority is experienced and perceived individually, addressing micro (newsroom-specific) tensions. Participatory engagement disrupts previously discussed dynamics that may encourage deference to AI-generated outputs, such as automation bias, anthropomorphism, and normative projection, by transforming the system from an external epistemic authority into a negotiated artifact \cite{simon_uneasy_2022, birhane_power_2022, suresh_participation_2024}. Importantly, this does not eliminate efficiency pressures or cognitive load, but systems that journalists have helped shape are more likely to be treated as provisional tools whose outputs remain open to contestation \cite{cooper_fitting_2024, bannon_reimagining_2018, bratteteig_disentangling_2012, inie_designing_2023}. 
Examples that move beyond traditional "weak" alignment span a spectrum of participatory depth, across four levels of participation, as per Delgado et al. \cite{delgado_participatory_2023}: consult, include, collaborate, and own. Consultative approaches, such as Constitutional AI \cite{bai_constitutional_2022}, and more recent Direct Preference Optimization (DPO) methods \cite{rafailov_direct_2024} incorporate human input primarily after core design choices have already been fixed. More inclusive practices expand this scope by integrating external expertise and plural perspectives into the development process, as in professional red-teaming, deliberative policy elicitation, and alignment datasets designed to preserve disagreement rather than collapse it into a single normative average \cite{ahmad_openais_2025, kirk_prism_2024, wu_honor_2023},  with frameworks such as Collective Constitutional AI, jury learning, and democratic policy development explicitly challenging majoritarian alignment \cite{huang_collective_2024, gordon_jury_2022, konya_democratic_2023}. At the collaborative end of the spectrum, "participatory ceilings" are broken when domain experts are acknowledged as co-creators of alignment targets, tools, and governance structures \cite{nayebare2023openai, lee_webuildai_2019}. Proposed alignment techniques such as Moral Graph Elicitation exemplify this shift by building on responsible AI design principles, granting participants delegated power over how values are formalized, reconciled, and translated into technical objectives, emphasizing transparency, contestability, and accountability \cite{klingefjord_what_2024, pushkarna_data_2022}. 
Finally, cases of collective ownership, (such as community-run language modeling efforts \cite{nekoto_participatory_2020}), further demonstrate how alignment can move from preference extraction toward durable forms of epistemic authority grounded in shared control, reframing AI systems as mediated artifacts shaped by affected users \cite{tseng_ownership_2025, corbett_power_2023}.

\subsection{Participatory Design as a Mechanism for Retaining Authority in Journalism} \label{participatory methods}

From an authority perspective, participatory methods matter because they reposition judgment. When journalists participate in defining training data boundaries, prompt templates, evaluation metrics, or newsroom-specific norms for acceptable outputs, they can intervene directly in the epistemic infrastructure that shapes AI integrations and, consequentially, the authority reconfigurations \cite{tseng_ownership_2025, ajmani_secondary_2025, suresh_participation_2024, cooper_fitting_2024, inie_designing_2023}. This shifts authority away from opaque, externally defined optimization practices and toward professional standards, retaining epistemic warrant within institutional contexts. Participation hence operates as a mechanism for reclaiming the externalized criteria through which journalistic knowledge claims are normally produced and validated when using commercial LLMs \cite{suresh_participation_2024}.

Importantly, participatory configurations build on the hybrid authority dynamics discussed above, making it legible and contestable. Beyond pseudo-transparency or generic XAI and human-in-the-loop narratives, participatory systems require early specification of roles, affiliations, and decision boundaries, rendering alignment auditable and institutionally legitimate \cite{hopkins_recourse_2025, balayn_unpacking_2025, corbett_power_2023, delgado_stakeholder_2021}. Hoque et al. \cite{hoque_towards_2024} demonstrate this in their framework for news-focused QnA chatbots, highlighting the importance of human oversight, clear authorial roles, and hybrid routing policies that distinguish factual from subjective content. In line with value-sensitive design criteria \cite{friedman_value_2003}, this emphasis on auditability and legitimacy ensures that contentious outcomes can be traced to documented editorial choices rather than absorbed into a moral buffer attributed to "the model". Responsibility is thus redistributed across explicit institutional processes, becoming collective but not diffuse, and anchored in recorded decisions. 
In practice, structured participation and co-design remain rare, but cases of emerging "trading zones" such as AI task forces illustrate hybrid roles \cite{xiao_it_2025,cools_uses_2024,european_broadcasting_union_leading_2025,albizu-rivas_artificial_2024}, bringing journalists and tech developers together to share responsibility and decision-making \cite{fischer_understanding_2011, thasler2025journalistic}. These examples, alongside tools and initiatives aimed at increasing journalists’ AI literacy \cite{annapureddy_generative_2025}, highlight that participatory AI in journalism is still largely exploratory, with limited systematic involvement or ownership among stakeholders. 

%limitations:

\subsection{Limitations of Participatory Authority } \label{limitations of participatory methods}

Participatory approaches should not be framed as a universal remedy, as participation can be selectively implemented, unevenly distributed, or reduced to pseudo-participation, where performative consultative practices are used to legitimize predetermined decisions rather than to meaningfully redistribute authority. In participatory AI, such dynamics often manifest as extractive arrangements in which participants function primarily as data sources or opinion validators, while core design choices remain centralized and constrained \cite{delgado_participatory_2023,sloane_participation_2022,palacin_design_2020,birhane_power_2022}.
Structural limitations further emerge when participation is expected to scale across complex and heterogeneous socio-technical systems. In this sense, the discussed notion of a "participatory ceiling" captures the limited capacity of situated actors to meaningfully influence AI systems designed as broadly applicable, a tension that is particularly salient in journalism, where contexts, roles, and values are highly variable \cite{delgado_stakeholder_2021, groves_going_2023, suresh_participation_2024}.
Participation is also shaped by persistent power asymmetries and material constraints. Technical experts and large institutions often retain control over problem formulation and final decision-making, while non-technical stakeholders are confined to influencing implementation rather than adoption \cite{ajmani_secondary_2025, cooper_fitting_2024, feffer_preference_2023, sorensen_value_2024}. Moreover, participatory processes demand time, expertise, and sustained engagement, imposing significant epistemic and labor burdens that are rarely compensated or institutionally supported. In such conditions, participation can increase individual workload while leaving structural authority unchanged \cite{rakova_where_2021, birhane_power_2022}.

For these reasons, \textbf{participatory authority retention}\textit{ }should be understood as \textit{conditional and situated rather than normative. Its effectiveness depends on where participation intervenes in the newsroom's socio-technical environment}: practices occurring early in the news-making processes, taking collective rather than individual forms, and being formally supported rather than improvised. When these conditions are absent, participatory gestures may coexist with, or even legitimize, the internal and external migrations of authority from journalists to AI tools and platforms.

\section{Discussion and Conclusion} \label{discussion}

Across the 209 records analyzed, discussions of AI adoption in journalism consistently converge on questions of \textit{who exercises editorial authority, whose knowledge claims receive epistemic legitimacy, and who bears responsibility for outcomes}. Although these concerns appear under different conceptual vocabularies across journalism studies, HCI, and FAccT research, our synthesis shows that they repeatedly manifest through two concurrent authority migrations (for an overview of representative quotes from key sources see Appendix \ref{Supplementary Material: Representative Discussion Quotes by Theme}).

Research on the digital transformation of newsrooms shows how editorial authority increasingly emerges from socio-technical assemblages in which computational systems participate in shaping editorial routines and decision environments \textbf{(Section \ref{digital transformation})} \cite{deuze2008changing, seaver_algorithms_2017, diakopoulos_automating_2019, thasler2025journalistic, zilberstein_models_2024}. Work on human–AI interaction further demonstrates how generative systems acquire perceived epistemic legitimacy through fluency, stylistic competence, and social cues aligned with professional journalistic practices \textbf{(Section \ref{perceived authority})} \cite{shin_automating_2025, smeenk_journalist_2023, norman_how_1994, reeves1996media, peter2025benefits, punday_2025_digital, akbulut2024all, zhou2023synthetic, nass_computers_1994, romeo2025exploring, carnat2024human}. Alignment practices and institutional voice mimicry further allow LLM outputs to reproduce journalistic norms and values, creating perceptions of institutional alignment that exceed the models’ normative grounding \textbf{(Section \ref{perceived authority})} \cite{khamassi_strong_2024, mazeika_utility_2025, komatsu2020ai, seaver_algorithms_2017, jakesch2023co, sundar2019machine}. At the same time, scholarship on oversight in AI-mediated workflows shows how these dynamics generate responsibility gaps in which journalists retain formal accountability while exercising reduced control over the epistemic processes producing content \textbf{(Section \ref{responsibility})} \cite{matthias2004resp, bomba_agency_2025, elish_moral_2019, tsamados_human_2025, schoeffer2025ai, dodds_controlled_2025}. Across these strands, scholars also document growing professional concern among journalists about the implications for editorial autonomy, accountability, and professional judgment (\ref{Supplementary Material: Concerns around Generative AI in Digital Journalism}). This literature ultimately points to an internal migration of editorial authority in which epistemic influence progressively shifts from journalists toward AI systems embedded in newsroom workflows.

Research on the political economy and organizational structure of digital media shows that editorial authority has long been shaped by dependencies on funding structures, infrastructures, ownership arrangements, and distribution systems, alongside hierarchical newsroom governance and contested editorial independence \textbf{(Section \ref{subsec:foundation_power})} \cite{breed_social_1955, pavlik_impact_2000, schudson_virtues_2005, jane_b_singer_contested_2007, karppinen_what_2016, humprecht_ownership_2019, ananny_2018_freedom, simon_uneasy_2022, drunen_safeguarding_2023}. Studies of platformization, algorithmic governance, and infrastructure provision show that AI adoption intensifies these dynamics by shifting decision-making power toward platforms, vendors, and technology companies that increasingly provide the tools, infrastructures, and distribution channels used in news production \textbf{(Section \ref{subsec:platform_mechanisms})} \cite{gillespie2014relevance, bell_platform_2017, nielsen2022power, chua2022platform, poell2023spaces, papaevangelou_funding_2024, simon_escape_2024}. Research on infrastructural capture and platform dependence further highlights how technical capabilities, datasets, and computational resources become concentrated among a small number of actors while external optimization logics are embedded into editorial processes \textbf{(Section \ref{subsec:platform_mechanisms})} \cite{aneesh2006virtual, jungherr2021digital, corbett_power_2023, hansen_et_al_initial_2023, simon_escape_2024, simon_artificial_2024, Christin2020, petre2015traffic}. These dependencies also reshape internal newsroom roles and decision structures, redistributing authority toward technical specialists, managers, and algorithmic systems while narrowing journalists’ autonomy and fragmenting accountability within digitized workflows \textbf{(Section \ref{subsec:internal_reconfigurations})} \cite{garcia_aviles_journalists_2004, anderson2013towards, diakopoulos2020computational, van_drunen_editorial_2021, kosterich2020managing, dodds2024impact, cools_news_2024, dorr2017ethical}. Collectively, this scholarship points to an external migration of editorial authority in which key capacities shaping news production and distribution increasingly reside in the infrastructures and technological providers on which journalism depends.

Participatory design and responsible AI research suggests that involving affected stakeholders in the development, alignment and implementation of AI systems in professional contexts such as journalism can reshape authority relations by redistributing decision rights and epistemic control, while different degrees of involvement create varying capacities for actors to influence model behavior, data selection, and evaluation criteria \textbf{(Section \ref{degrees of participation})} \cite{arnstein_ladder_1969, delgado_participatory_2023, suresh_participation_2024, birhane_power_2022, delgado_stakeholder_2021, ajmani_secondary_2025, tseng_ownership_2025, inie_designing_2023}. Such approaches can help retain editorial authority in newsrooms by making the epistemic infrastructure of AI systems more transparent, contestable, and institutionally grounded \textbf{(Section \ref{participatory methods})} \cite{suresh_participation_2024, cooper_fitting_2024, hoque_towards_2024, corbett_power_2023}. However, participation remains conditional: structural power asymmetries, limited resources, and risks of pseudo-participation can leave underlying authority configurations largely unchanged or even legitimized \textbf{(Section \ref{limitations of participatory methods})} \cite{sloane_participation_2022, palacin_design_2020, rakova_where_2021, birhane_power_2022}.

Taken together, these findings indicate that \textbf{fairness, accountability, and transparency in AI-mediated journalism cannot be addressed solely through improvements in model performance or explainability}. Instead, effective FAccT interventions must engage directly with the \textbf{authority reconfigurations} that structure how AI systems are integrated into newsroom practices and infrastructures, recognizing that \textbf{editorial authority should be treated not as an outcome of successful AI adoption but as a necessary precondition for FAccT-compliant implementations}. From this perspective, future research should move beyond purely technical evaluations of AI and examine how these technologies reshape editorial roles, responsibilities, and decision-making structures within news organizations. We therefore call for in-depth empirical studies that trace how authority reconfigurations vary across organizational contexts, degrees of journalists’ participation in AI development and deployment, and different newsroom environments, treating AI not only as a tool but as an infrastructural actor embedded in broader institutional and professional dynamics.

\section* {Acknowledgments}
This work was funded by the EU Horizon Europe program, Marie Skłodowska-Curie Actions (MSCA), alignAI project (grant  101169473).
We thank the reviewers for their time and valuable feedback. We also thank our colleagues at the Social Computing Group for their feedback and support throughout this work.

\section* {Generative AI Disclosure Statement}
In the preparation of this manuscript, large language models were exclusively used for formatting assistance and language support, including arranging table contents, refining paragraph layout and spacing, and performing proofreading, spelling, grammar, and lexical checks. All substantive content, analysis, and interpretations were conceived and written by the authors.

\bibliographystyle{ACM-Reference-Format}
\bibliography{bibliography.bib}

\clearpage  % Force a page break

\appendix
\section{Supplementary Material: Concerns around Generative AI in Digital Journalism}
\label{Supplementary Material: Concerns around Generative AI in Digital Journalism}

\begin{longtable}{p{0.45\linewidth} p{0.45\linewidth}}
\caption{Concerns around Generative AI in Digital Journalism, with corresponding references}\\
\toprule
\textbf{Concerns} & \textbf{Sources}\\
\midrule
\endhead
Accuracy and Reliability (including Hallucinations, Misinformation, Fictitious Citations, Wrong Information, Unreliability) & \cite{amponsah2024navigating, becker_policies_2025, beckett2023generating, cools_towards_2023, cools_news_2024, diakopoulos_generative_2024, hoque_towards_2024, jones2023generative, longoni_news_2021, moran_robots_2022, moller_reinforce_2024, nishal_envisioning_2024, tseng_ownership_2025, wang_reelframer_2024, john_wihbey_ai_2024, xiao_it_2025, van_dalen_revisiting_2024}\\
\addlinespace
Bias (including Algorithmic Bias, Reinforcing Stereotypes, Undermining Editorial Values) & \cite{amponsah2024navigating, becker_policies_2025, cools_towards_2023, cools_uses_2024, jones2023generative, nishal_envisioning_2024, sonni_digital_2024, tseng_ownership_2025}\\
\addlinespace
Job Displacement and Changes to Journalistic Labor/Identity/Professional Status & \cite{amponsah2024navigating, diakopoulos_generative_2024, moran_robots_2022, moller_reinforce_2024, simon_artificial_2024, tseng_ownership_2025, van_dalen_revisiting_2024}\\
\addlinespace
Transparency and Accountability & \cite{amponsah2024navigating, becker_policies_2025, diakopoulos_generative_2024, cools_news_2024, dodds2024impact, hoque_towards_2024, jones2023generative, moran_robots_2022, nishal_envisioning_2024, oh2025harmonizing, paik_journalism_2025, tseng_ownership_2025, john_wihbey_ai_2024}\\
\addlinespace
Editorial Autonomy and Human Control/Oversight & \cite{cools_towards_2023, cools_uses_2024, tseng_ownership_2025, xiao_it_2025}\\
\addlinespace
Influence and Power of Technology Companies/Platform Dependency & \cite{becker_policies_2025, simon_artificial_2024}\\
\addlinespace
Audience Trust and Perception & \cite{dodds2024impact, hoque_towards_2024, longoni_news_2021, nishal_envisioning_2024, tseng_ownership_2025, john_wihbey_ai_2024, wu_journalists_2024, van_dalen_revisiting_2024}\\
\addlinespace
Ethical Frameworks and Guidelines (including Ethical Boundaries, Journalistic Norms and Values) & \cite{amponsah2024navigating, cools_news_2024, diakopoulos_generative_2024, hoque_towards_2024, jones2023generative, moller_reinforce_2024, nishal_envisioning_2024, oh2025harmonizing, paik_journalism_2025, sonni_digital_2024, tseng_ownership_2025, xiao_it_2025, van_dalen_revisiting_2024}\\
\addlinespace
Intellectual Property and Data Privacy (including Copyright, Data Training) & \cite{beckett2023generating, diakopoulos_generative_2024, simon_artificial_2024, sonni_digital_2024, tseng_ownership_2025}\\
\addlinespace
Lack of Originality/Homogenized Content & \cite{diakopoulos_generative_2024}\\
\addlinespace
Cost and Resource Constraints & \cite{nishal_envisioning_2024}\\
\addlinespace
AI Literacy and Understanding & \cite{cools_uses_2024}\\
\addlinespace
Regulatory Challenges & \cite{cools_uses_2024, oh2025harmonizing}\\
\addlinespace
Risk of ``Epistemic Capture''/``Lock-in'' & \cite{john_wihbey_ai_2024}\\
\addlinespace
Uncertainties and Threats impacting Willingness to Use AI & \cite{wu_journalists_2024}
\\
\bottomrule
\end{longtable}

\section{Supplementary Material: AI Journalism Atlas - Complete Overview}
\label{Supplementary Material: AI Journalism Atlas - Complete Overview}
This appendix provides a comprehensive mapping of artificial intelligence functionalities currently deployed or under development across the journalistic workflow, from newsgathering through production to distribution. The atlas synthesizes academic characterizations of AI capabilities with documented evidence of real-world newsroom implementations, serving two purposes: first, it establishes the scope and diversity of AI integration in contemporary journalism; second, it provides concrete grounding for the authority reconfigurations analyzed in the main text. Each entry identifies the workflow stage where AI intervenes, describes the functionality and its operational mechanisms as characterized in academic literature, assesses the technology's current deployment status (e.g., implemented, experimental, prototype), cites relevant academic sources, and documents concrete examples from research as well as named news organizations. Operational descriptions and status assessments derive from the cited academic sources; concrete examples (with hyperlinks to the sources) illustrate reported real-world deployments. This mapping reveals how AI adoption spans the entire news production chain, from initial story discovery to final audience delivery, creating multiple points where editorial authority can migrate from human journalists to algorithmic systems or from newsrooms to external technology providers.

\begin{longtable}
{p{0.15\linewidth}p{0.25\linewidth}p{0.20\linewidth}p{0.10\linewidth}p{0.20\linewidth}}
\caption{Atlas of AI functionalities across the journalistic workflow. Operational descriptions and status assessments are derived from academic literature cited in the "Academic Sources" column; concrete examples document reported research and real-world newsroom deployments.}\\
\toprule
\textbf{Functionality/ Workflow Stage} & \textbf{Operational Description} & \textbf{Status} & \textbf{Academic Sources} & \textbf{Concrete Examples (with hyperlinks)}\\
\midrule
\endhead
Content Discovery and Sense-making/ Newsgathering & LLMs perform extractive and abstractive summarization of complex documents to identify newsworthy angles, querying documents, requesting quotes, or brainstorming counter-arguments to claims. & Hybrid: Traditional supervised learning and NLP tools successfully piloted and used in newsrooms. Generative AI applications currently exploratory, though recent advances show significant promise. & \cite{nishal_envisioning_2024} & \href{https://digiday.com/media/inside-reuters-agentic-ai-video-experiment/}{Reuters (Fact Genie)}\\
\midrule
 Newsworthiness Ranking/ Newsgathering & Algorithmic support systems and supervised learning perform statistical evaluations and automated detection of anomalous/interesting phenomena, identifying newsworthy leads by comparing them against predictive models of news value. & Implemented/Piloted: Traditional supervised learning and NLP technologies successfully piloted and used in newsrooms. & \cite{nishal_envisioning_2024} & Reuters Tracer \cite{liu2016reuters}; \href{https://www.dataminr.com/products/dataminr-for-news/}{Dataminr for News}, \href{https://www.reuters.com/article/rpb-cyber/the-cybernetic-newsroom-horses-and-cars-idUSKCN1GO0Z0/}{Reuters (Lynx Insights)}\\
\midrule
 Data Mining and Analytics/ Newsgathering & Systems perform data triangulation and automated reading and summarization of documents by scraping, analyzing, and processing massive amounts of available data to identify patterns. In investigative contexts, includes checking data for contempt of court risks or ethically vetting source disclosures. & Rapidly Evolving / Risk-Assessment Stage: AI applications in news have grown in recent years but environment remains rapidly changing with slow legal response. Industry currently lacks advanced understanding of system failures, undergoing rapid risk-based review before widespread safe integration. & \cite{jones2023generative} & \href{https://www.ire.org/2025/12/06/inside-the-new-york-timess-a-i-toolkit/}{New York Times (Cheat Sheet)}, \href{https://www.axios.com/2024/08/20/washington-post-ai-tool-data}{Washington Post (Haystacker)}, \href{https://source.opennews.org/articles/linked-data-bbc/}{BBC News Labs (Juicer)}, \href{https://qz.com/1670632/how-quartz-used-ai-to-help-reporters-search-the-mauritius-leaks}{Quartz/ICIJ (doc2vec-Mauritius Leaks)}\\
\midrule
 Automated source monitoring/ Newsgathering & The use of AI systems to scan vast information ecosystems for monitoring and reviewing information from materials, events, or the environment in order to detect or assess problems. & Rapidly Changing / High-Risk Transition: The news industry currently lacks an advanced understanding of exactly how these systems work or why they fail. The current era is defined by a rush to automate and a competitive ‘arms race’ among technology providers, which leads to necessary checks and balances being ignored or deprioritized in favor of establishing market dominance & \cite{jones2023generative} & \href{https://www.newswhip.com/2016/04/guest-post-associated-press-use-newswhip/}{Associated Press (NewsWhip)}, \href{https://www.cjr.org/tow_center/an-algorithmic-nose-for-news.php}{Duke Tech \& Check Cooperative (ClaimBuster monitoring CNN)}\\
\midrule
 Data scraping and extraction/ Newsgathering and Information Gathering) & AI performs technical tasks such as manipulating spreadsheets, small data cleanup tasks, and extracting information from large document corpora to reduce repetitive tasks and save time by transforming unstructured documents into formats easier for journalists to analyze. & Experimental / Aspirational: While 7.7\% of respondents currently work with data using generative AI, it is primarily identified as high-priority area for aspirational usage. Journalists view it as unmet opportunity due to tasks being highly tedious, boring, or repetitive. & \cite{diakopoulos_generative_2024} & \href{https://www.ire.org/2025/12/06/inside-the-new-york-timess-a-i-toolkit/}{New York Times (Cheat Sheet)}, \href{https://www.themarshallproject.org/2023/10/03/banned-books-prisons-how-to-report?}{Marshall Project (AI-assisted extraction/summarization workflow)}, \href{https://digiday.com/media/inside-reuters-agentic-ai-video-experiment/}{Reuters (Fact Genie)}\\
\midrule
 Data verification support/ Newsgathering and Production & AI tools automate detection of anomalies and verification of data sets. Systems serve as first check rather than final authority; journalists use them to flag potential issues but rely on human judgment and journalistic expertise for complex decisions. & Available but under-utilized: Technological capability exists and is used by some, but source notes low adoption rate and cultural resistance. Journalists often prefer familiar tools like Google due to high learning curves of in-house verification tools or lack of systemic blend with established editorial routines. & \cite{gutierrez_lopez_question_2023} & \href{https://www.invid-project.eu/tools-and-services/invid-verification-plugin/}{InVID-WeVerify Verification Plugin}, \href{https://innovation.dw.com/articles/truly-media-beta-has-crossed-the-finishing-line?}{Deutsche Welle (DW)—Truly Media collaborative verification platform}, \href{https://mevanbabakar.com/2018/09/27/citizen-beta-statistics-factchecking-and-data-journalism/}{Full Fact (UK)—"Robochecking" against ONS data/API}, \href{https://www.rand.org/research/projects/truth-decay/fighting-disinformation/search/items/claimbuster.html}{ClaimBuster}, \href{https://cordis.europa.eu/project/id/101070093?}{vera.ai EU (multimodal disinformation detection)}. Note: Limited widespread adoption in newsrooms.\\
\midrule
 Sentiment analysis of UGC/ Interrelated (Newsgathering, Production, and Distribution) & NLP and Neural Networks (often via third-party tools like Google Cloud APIs) process millions of news items or social media posts. Machine learning models classify content (e.g., good or bad comments) to reduce toxicity and promote engaging interactions. & Implemented / Aspirational: Basic versions such as Natural Language sentiment and comment moderation tools used in newsrooms for several years. Deeper sentiment analysis on UGC categorized as top ten ways newsrooms expect AI to reshape journalism in future. & \cite{beckett2019new} & \href{https://digiday.com/media/new-york-times-moderates-12000-comments-day/}{New York Times Moderator}, \href{https://www.wired.com/2017/02/googles-troll-fighting-ai-now-belongs-world/}{Guardian and The Economist (Perspective/Jigsaw API experiments)}\\
\midrule
 Automated Story Generation (Automated Journalism)/ News Production & Algorithms process structured datasets (financial reports, sports scores, weather patterns, election data) to generate text, video, or audio narratives. Earlier systems relied on template-based approaches; modern iterations increasingly use LLMs to produce complex story drafts and personalized content. & Implemented / Rapidly Evolving: 73\% of news organizations report using AI for news writing automation. Basic commodity journalism institutionalized and produces thousands of stories across regions. Shift toward generative AI for autonomous content currently in high-stakes arms race phase, characterized by rapid development and focus on managing factual accuracy and ethical risks. & \cite{sonni_digital_2024, beckett2019new, jones2023generative, cools_news_2024, lewis_et_al_pdf_2019} & \href{https://www.ilfoglio.it/il-foglio-ai/}{Il Foglio AI}, \href{https://www.intotheminds.com/blog/en/yle-voitto-robot-artificial-intelligence/}{Voitto robot}, \href{https://www.washingtonpost.com/pr/wp/2016/10/19/the-washington-post-uses-artificial-intelligence-to-cover-nearly-500-races-on-election-day/}{Washington Post (Heliograf)}, \href{https://www.ap.org/solutions/artificial-intelligence/}{Associated Press (Wordsmith)}, \href{https://www.nytimes.com/2019/02/05/business/media/artificial-intelligence-journalism-robots.html}{Bloomberg (Cyborg)}, \href{https://www.reuters.com/article/rpb-cyber/the-cybernetic-newsroom-horses-and-cars-idUSKCN1GO0Z0/}{Reuters (Lynx Insights)}\\
\midrule
 Headline Generation/ News Production & Generative AI and LLMs create news headlines based on provided data or article drafts. Journalists use these tools as collaborator or sounding board to iterate on ideas. Research indicates significant algorithm aversion among audiences: AI-generated headlines perceived as less accurate and result in lower trust than those attributed to human reporters. & Implemented / Regular Usage: Dominant category of current use in industry. Leading media organizations including Associated Press, The New York Times, and The Washington Post already use AI to assist in generating content at headline level. & \cite{diakopoulos_generative_2024, longoni_news_2021} & \href{https://www.forbes.com/sites/forbesproductgroup/2018/07/11/entering-the-next-century-with-a-new-forbes-experience/}{Forbes (Bertie CMS)}, \href{https://www.zenml.io/llmops-database/global-news-organization-s-ai-powered-content-production-and-verification-system}{Reuters (LEON CMS)}, \href{https://www.ap.org/solutions/artificial-intelligence/}{Associated Press (AI headline generation)}\\
\midrule
 Abstractive Summarization and Question Answering/ Newsgathering and News Production & LLMs process unstructured datasets such as legislative documents, consumer reports, and jargon-heavy technical papers. Journalists and users interact with models through chatbot-style querying to interrogate data and ask for specific angles or use pre-engineered prompts to generate bulleted briefs or story drafts. Major operational risk is hallucination, where system generates plausible-sounding but factually incorrect information not found in source text. & Implemented / Regular Usage: Summarization and Question-Answering is one of dominant categories of current generative AI use in newsrooms. Performance rapidly improving, currently positioned as tool to kickstart writing process or support rapid discovery, always requiring human oversight to verify accuracy. & \cite{nishal_envisioning_2024, diakopoulos_generative_2024} & \href{https://www.ire.org/2025/12/06/inside-the-new-york-timess-a-i-toolkit/}{New York Times—Echo}, \href{https://www.washingtonpost.com/ask-the-post-ai/}{Washington Post—Ask The Post AI}, \href{https://ask.ft.com/signed-out?signin-url=https://accounts.ft.com/login?location=https://ask.ft.com/}{Financial Times—Ask FT}, \href{https://www.forbes.com/search/}{Forbes — Adelaide}, \href{https://www.ap.org/solutions/artificial-intelligence/}{Associated Press (Minutes Project for public meeting transcription and summarization)}\\
\midrule
 Technical document simplification/ Newsgathering and News Production & LLMs perform abstractive summarization and simplification specifically on unstructured datasets containing technical data such as legislative and policy documents or consumer reports. Journalists use crafted prompts to specify desired features such as higher-level simplicity or bulleted format. Process requires human oversight to ensure simplified text remains factually accurate and free from hallucinations. & Emerging / Proposed Augmentation: Source identifies this as area where performance is continuing to improve and presents significant promise for augmenting work, though currently noted as being under review in specialized journalistic contexts. & \cite{nishal_envisioning_2024} & \href{https://www.themarshallproject.org/2023/03/16/prison-banned-books-policies-what-we-know?}{The Marshall Project}. Note: Limited verified newsroom examples.\\
\midrule
 News angle generation/ Newsgathering & LLMs supplement sense-making from unstructured datasets through chatbot-style interactions for brainstorming or through pre-engineered prompts designed to evaluate text through lens of specific news values (e.g., identifying if document contains something controversial, surprising, or novel). & Emerging / Proposed: While showing significant promise in recent research and pilot tools, systems currently viewed as way to supplement rather than substitute existing human routines of news discovery. & \cite{nishal_envisioning_2024} & AngleKindling \cite{petridis2023anglek}, \href{https://rm.coe.int/ai-in-the-media-en-final-publish/488029acb3}{RTS BakerStreet}\\
\midrule
 Automated graphic generation/ News Production & AI tools speed up news production procedures and optimize real-time coverage by automating creation of visual assets. Tools intended to release journalists from menial tasks, allowing focus on in-depth investigations requiring human expertise. Tools must be designed to blend with editorial requirements as human judgment remains necessary for complex decisions. & Implementation Phase / Gaining Momentum: Use of emerging technologies in journalism has gained momentum but often faces cultural resistance and struggle for full newsroom adoption. Some journalists upbeat about opportunities while others wary of mistakes or being replaced, making sociotechnical design and close collaboration between journalists and technologists critical for successful embedding. & \cite{gutierrez_lopez_question_2023} & \href{https://wan-ifra.org/2021/06/how-bbc-news-labs-uses-ai-powered-content-automation-to-engage-young-audiences/}{BBC News Labs (Graphical storytelling)}\\
\midrule
 Visual article illustration/ News Production and Content Production & Journalists use written prompts to generate numerous visual iterations of a concept rapidly, allowing newsrooms to explore creative options and styles previously restricted by time or budget. Functionality often integrated into existing software such as Adobe Photoshop's generative AI tools, which graphics teams use to save hours of time each day. Tools serve as creativity support to kickstart visual elements of a story. & Implemented / Regularly Used: Approximately 20.4\% of news practitioners report using generative AI for multimedia production (including illustrations) on experimental or regular basis. High-priority area for aspirational use (23.1\%), with organizations like Financial Times already exploring AI-augmented visuals while emphasizing need for clear disclosure to reader. & \cite{nishal_envisioning_2024, diakopoulos_generative_2024} & \href{https://www.ilfoglio.it/il-foglio-ai/}{Il Foglio AI-AI generated Images}, \href{https://archive.ph/0vRTz}{The Economist (Midjourney for cover's design)}, \href{https://lampoonmagazine.com/vanni-santoni-ai-arte-prompting-fumetti/}{Corriere della Sera Italy (Midjourney generated comic, August 2022)}, \href{https://www.theatlantic.com/newsletters/archive/2022/08/where-does-alex-jones-go-from-here/676843/}{The Atlantic (Midjourney for Alex Jones newsletter images)}.\\
\midrule
 Automated video production and AI anchors/ News Production & Generative AI and audio-visual deepfake technology produce persuasive and compelling video content. Systems automatically create news videos or deploy synthetic media to deliver narratives, often designed to offer new, innovative forms of news presentation. & Emerging / Innovative: While organizations increasingly turning to video automation to scale production, use of deepfake-style synthetic media considered new dimension of possibilities that currently faces challenges related to transparency and risk of media manipulation. & \cite{cools_news_2024, sonni_digital_2024} & \href{https://www.scmp.com/tech/innovation/article/2172235/xinhua-news-agency-debuts-ai-anchors-partnership-search-engine-sogou}{Xinhua–Sogou AI anchors}, \href{https://koreajoongangdaily.joins.com/2020/11/10/entertainment/television/MBN-AI-artificial-intelligence/20201110153900457.html}{MBN AI news anchor}, \href{https://www.adweek.com/tvspy/has-it-begun-ai-anchor-starts-working-at-station-in-india/}{Aaj Tak/India Today Group AI anchor}, \href{https://www.heygen.com/customer-stories/studio-47}{STUDIO 47 (Germany)—HeyGen avatars}, \href{https://www.theweathernetwork.com/en/news/weather/forecasts/the-weather-network-launches-first-ever-ai-assisted-avatar?}{The Weather Network (Canada)—AI avatars}, \href{https://www.theguardian.com/world/article/2024/aug/27/venezuela-journalists-nicolas-maduro-artificial-intelligence-media-election}{Connectas Venezuela-AI avatars}\\
\midrule
 Storyboard generation/ News Production & Human-AI collaborative workflow divided into two main panels. First, character board is created where LLM suggests descriptions for costumes, props, and backgrounds to distinguish characters, then rendered by text-to-image model. Second, system generates shot descriptions (specific instructions for facial expressions, gestures, and actions) for every line of dialogue. Elements synthesized into visual frames, often rendered in black-and-white line art to mimic traditional storyboarding styles. Users can edit, regenerate, or rewrite descriptions at any step to ensure visual plan is accurate and achievable for human actor to film. & Experimental / Prototype: Currently implemented in research systems like ReelFramer, used to scaffold process for journalists learning to create short-form social media content. While effective for ideation, technology faces challenges regarding visual consistency across different frames and accurately portraying multiple specific characters in single shot. & \cite{wang_reelframer_2024} & \href{https://tech.ebu.ch/news/2025/03/generative-ai-promising-for-content-creation-but-not-yet-ready-for-primetime}{EBU technology project in collaboration with RAI (script + concept art + storyboard for a pilot)}.\\
\midrule
 Automated transcription/ Newsgathering and News Production & Speech-to-text algorithms process audio files at scale, acting as productivity tool relieving journalists from more time-consuming task of manual transcription. While highly effective for English, current systems sometimes struggle with other languages or specific dialects, leading newsrooms to train algorithms with specialized dialect data to improve accuracy. & Implemented / Regularly Used: Functionality widely adopted and viewed as clear-cut success for increasing newsroom efficiency. However, better automated transcriptions remains one of top ten ways newsrooms expect AI to continue reshaping industry in future. & \cite{nishal_envisioning_2024, beckett2019new, moran_robots_2022} & \href{https://trint.com/blog/trint-launches-instant-transcription-tool}{Trint-beta tested at Thomson Reuters, BBC, NPR and The Washington Post}, \href{https://www.ap.org/the-definitive-source/products-and-services/introducing-5-ai-solutions-for-local-news/}{Associated Press Local News AI Initiative}, \href{https://reutersagency.com/media-centre/reuters-launches-ai-powered-discoverability-features-for-video-library-on-reuters-connect-accelerating-discovery-editing-and-publishing?}{Reuters-automated transcripts, translation and identification for video}\\
\midrule
 Machine Translation/ News Production & LLMs and specialized tools to automatically convert articles, scripts, or social media posts into multiple languages, allowing newsrooms to reach global audiences and underrepresented language communities more efficiently, sometimes replacing need for human freelancers for basic translation tasks. Significant risk is linguistic inequality as generative models often perform significantly better in dominant Western languages than others, impacting accuracy of information for certain global populations. & Implemented / Rapidly Improving: Performance continuing to improve, currently used on regular or experimental basis by 8.8\% of news practitioners. While already established as productivity tool, expected to eventually become invisible part of standard journalistic workflow. & \cite{nishal_envisioning_2024, diakopoulos_generative_2024} & \href{https://assets.ctfassets.net/pplyeawnfzc7/5FXYD7j4zmA9iVqCyBgQ87/440a15637b681a480d3dac8ca3749fdd/Case_study-Nikkei_EN.pdf}{Nikkei Japan (DeepL Pro - translates for Nikkei Asia/FT)}, \href{https://reutersagency.com/content/content-types/ai-transcripts/}{Reuters (AVISTA - transcribes and translates speeches with time-coded results)}, \href{https://www.bbc.co.uk/rdnewslabs/news/multilingual-assessment/}{BBC News Labs (extensive testing of DeepL, Google, AWS, Azure, Whisper)}, \href{https://latamjournalismreview.org/articles/inside-a-puerto-rican-newsrooms-experiment-with-ai-powered-translations-to-reach-english-speaking-audiences/}{Centro de Periodismo Investigativo Puerto Rico (custom ChatGPT for Spanish→English)}\\
\midrule
 Automated tagging/ News Production & AI tools automatically assign filters and tags to journalistic content, allowing sped-up production procedures and optimized real-time coverage. Systems intended to release journalists from menial tasks, enabling time redirection toward in-depth investigations and high-value reporting. For successful adoption, tools must be developed using sociotechnical design that blends technological capabilities with specific editorial requirements and journalistic values. & Implementation Phase / Gaining Momentum: While technologies increasingly pervasive in newsrooms, often face cultural resistance from staff who worry about algorithmic mistakes or job replacement. Successful embedding relies on close collaboration between journalists and technologists to ensure tools fit existing work routines. & \cite{gutierrez_lopez_question_2023} & \href{https://www.bbc.co.uk/webarchive/https%3A%2F%2Fwww.bbc.co.uk%2Fblogs%2Finternet%2Fentries%2Faf2ffbc4-1758-3697-bb6b-ef4996b17b46}{BBC (Juicer with DBpedia - automatic semantic tagging in 4 categories: people, places, organizations, things)}, \href{https://www.reuters.com/media-center/reuters-launches-ai-powered-discoverability-features-for-video-library-on-reuters-connect-accelerating-discovery-editing-and-publishing/}{Reuters (AVISTA - facial recognition, shot change detection, automatic metadata)}, \href{https://nytlabs.com/projects/editor.html}{New York Times (Editor experimental 2015 - journalists highlight phrases, computer learns tags)}\\
\midrule
 Creative Support Tools/ Newsgathering and News Production & LLMs act as collaborator or sounding board to bounce off ideas and help journalists move past blank page. In newsgathering, support ideation and sense-making through chatbot-style querying to identify newsworthy angles or news values. In production, generate outlines, briefs, and first drafts to kickstart writing, and offer iterative editing by suggesting improvements for fluency, coherence, and jargon reduction. Primary risk is automation bias where authoritative-sounding outputs lead to over-reliance, and cliched recommendations that may devalue reporter's original voice. & Implemented / High Aspirational Interest: Currently approximately 21.5\% of practitioners use GenAI for information gathering and ideation. Major area of aspirational growth, with 34.9\% of news workers wanting to use it more for news discovery and research. & \cite{nishal_envisioning_2024, diakopoulos_generative_2024} & \href{https://www.forbes.com/sites/forbesproductgroup/2018/07/11/entering-the-next-century-with-a-new-forbes-experience/}{Forbes (Bertie - thought starter recommending trending topics, synthesizing research, learning writer style)}, \href{https://digiday.com/media/inside-reuters-agentic-ai-video-experiment/}{Reuters (internal sandbox with chain prompts; CMS Draft Tool processing press releases + interview transcripts)}, \href{https://www.ap.org/news-highlights/best-of-the-week/honorable-mention/2025/ap-uses-ai-to-pore-through-podcasts-for-revealing-story/}{Associated Press (Podcast Monitoring Tool - scrapes RSS, transcribes, identifies keywords; tracked 100+ Kash Patel FBI nomination podcast appearances)}\\
\midrule
 Deepfake detection and synthetic media recognition/ Newsgathering and News Production & Machine learning tools detect ML fakes by analyzing media for signs of sophisticated manipulation (audio-visual deepfakes) to help newsrooms verify authenticity in information environment where line between genuine and manipulated content increasingly blurred. Process vital for maintaining journalistic integrity and preventing inadvertent publication of fabricated information. & Emerging / High-Priority Growth Area: Identified as key growth area for newsrooms and one of top ten ways AI expected to reshape industry. Research in using AI to fight disinformation has seen 56\% increase since 2021, reflecting rising importance in digital transformation of newsrooms. & \cite{beckett2019new, sonni_digital_2024} & \href{https://www.afp.com/en/medialab}{AFP Medialab (InVID-WeVerify browser extension)}, \href{https://www.truemedia.org/post/free-deepfake-detector-to-help-newsrooms-across-the-country?}{TrueMedia (Free AI-enabled Deepfake Detector)}\\
\midrule
 AI-assisted debunking/ Newsgathering and Verification & Systems employ claim detection using machine learning to separate actual claims from other sentences to help human fact-checkers prioritize what to check. Includes claim matching, which identifies previously checked claims appearing in new contexts, and robochecking, which automatically verifies claims against reliable databases to provide real-time answers and evidence-based graphics. & Implemented / High Growth Area: Currently viewed as clear-cut success for newsroom efficiency, remains high-priority area as newsrooms seek to counter misinformation operating at scale. Research in using AI to fight disinformation has seen 56\% increase since 2021, highlighting critical role in modern digital transformation. & \cite{beckett2019new, sonni_digital_2024} & \href{https://fullfact.org/blog/2020/jul/afc-global/}{Full Fact UK (Full Fact AI, BERT-based)}, \href{https://chequeado.com/como-los-chequeadores-podemos-aprovechar-el-impulso-de-la-inteligencia-artificial-generativa/}{Chequeado Argentina (Chequeabot)}, \href{https://africacheck.org/}{Africa Check (Full Fact AI tools)}, \href{https://bassconnections.duke.edu/project/data-and-technology-fact-checking-2018-2019/}{Duke/UT Arlington (Squash/ClaimBuster)}\\
\midrule
 Personalized content distribution/ News Distribution & AI systems utilize real-time data and metrics (engagement, shares, time on page) to inform content strategies and optimize newsfeeds. Systems use collaborative filtering and machine learning to predict user propensity to subscribe or engage. Marks shift from traditional editorial logic to algorithmic logic where machine decides visibility based on user affinity, popularity, and timing. Creates bidirectional interactive feedback loop but risks creating filter bubbles or confirmation bias by hiding content algorithm deems uninteresting, potentially limiting exposure to diverse viewpoints. & Implemented / Pervasive: Over 62\% of news organizations currently use content personalization. Mature technology among digital intermediaries but continues to evolve in newsrooms through smarter paygates and personalized news narratives. Approximately half of news practitioners surveyed in early reports already employing AI for distribution. & \cite{beckett2019new, jones2023generative, cools_news_2024, outi_lundahl_algorithmic_2022} & \href{https://www.niemanlab.org/2018/02/after-years-of-testing-the-wall-street-journal-has-built-a-paywall-that-bends-to-the-individual-reader/}{Wall Street Journal (Dynamic Paywall)}, \href{https://digiday.com/media/news-uk-cut-subscriber-churn-half-using-self-learning-algorithm/}{The Times and The Sunday Times (JAMES-digital butler personalizing newsletters)}, \href{https://knightlab.northwestern.edu/2015/06/03/how-the-washington-posts-clavis-tool-helps-to-make-news-personal/}{Washington Post (Clavis)}, \href{https://knightlab.northwestern.edu/2016/03/28/a-quick-look-at-recommendation-engines-and-how-the-new-york-times-makes-recommendations/}{New York Times (Hybrid LDA Recommender)}\\
\midrule
 Social media curation and filtering/ News Distribution & Platforms replace traditional human editorial judgment with algorithmic logic utilizing real-time user data (likes, shares, browsing history) to optimize for engagement. Systems create personalized feeds by prioritizing content user likely to find personally informative or consistent with existing beliefs. Process risks prioritizing virality and shareability over journalistic quality, often resulting in algorithmic promotion of sensationalism or misinformation over high-civic-value reporting. As gatekeepers, systems can create filter bubbles by down-ranking diverse viewpoints or complex news that doesn't trigger immediate engagement. & Implemented / Dominant: Social media now primary news source for majority of population (62\% in U.S.), making these algorithms most influential editors in world. Status currently highly dynamic and under scrutiny; platforms frequently iterate algorithms (such as shifting priority toward friends and family) to address public outcries over fake news and pollution of information ecosystem. & \cite{jones2023generative, bell_platform_2017} & \href{https://www.niemanlab.org/2015/08/the-new-york-times-built-a-slack-bot-to-help-decide-which-stories-to-post-to-social-media/}{New York Times (Blossom Slack bot)}\\
\midrule
 Chatbots and voice agents/ News Production and Distribution & Systems utilize generative AI and LLMs to simulate human-like dialogue, allowing users to query archives or receive automated news updates via messaging platforms like WhatsApp. Voice agents optimize content for new mediums by converting text stories into audio formats for hands-free consumption. Offer efficiency in handling queries but carry risks of hallucinations (fictional data), automation bias, and potential disintermediation where users might prompt chatbot for news instead of visiting news organization's primary site. & Implemented / Rapidly Expanding: Approximately 85\% of newsrooms have at least experimented with generative AI, and 80\% expect larger future role for AI in operations. While basic chatbots for notifications are common, deep conversational systems that act as interface to all of publisher's content remain high-priority aspirational goal. & \cite{jones2023generative, diakopoulos_generative_2024, sonni_digital_2024, beckett2023generating} & \href{https://maldita.es/nosotros/20210603/disinformation-whatsapp-chatbot-frequently-forwarded-attribute/}{Maldita.es Spain (WhatsApp Chatbot)}, \href{https://www.aosfatos.org/noticias/ajude-a-testar-a-fatimagpt-nova-versao-da-robo-checadora-do-aos-fatos/}{Aos Fatos Brazil (FátimaGPT)}, \href{https://www.turn.io/news/ifcn-chatbot?}{IFCN Global (WhatsApp Chatbot)}, \href{https://factchequeado.com/institucional/20250312/factchequeado-chatmigrante-immigration-information/}{Factchequeado (ChatMigrante)}\\
\midrule
 Paywall optimization/ News Distribution and Business Operations & AI systems utilize machine learning algorithms to analyze real-time user metrics (engagement, frequency of visits, time on page) to predict specific user's propensity to subscribe or likelihood of churning (canceling). Wall Street Journal's dynamic paywall uses these models to show different visitors different levels of access based on predicted likelihood of signing up. Smarter, dynamic paygates move away from static rules to increase conversion rates and optimize revenue. Significant societal risk is potential for digital divide where high-quality news becomes restricted to those who can afford it while others left with unreliable or low-quality information. & Implemented / Core Industry Trend: Many large legacy organizations (The Washington Post, The Times of London) have already integrated AI into paywall operations. Currently ranked as one of top ten ways newsrooms expect AI to reshape industry through dynamic pricing for subscriptions. & \cite{jones2023generative} & \href{https://www.zuora.com/our-customers/case-studies/the-financial-times/}{Financial Times (AI Paywall with Zuora)}, \href{https://digiday.com/media/forbes-launches-dynamic-ai-paywall-as-it-ramps-up-post-search-commercial-diversification-plans/}{Forbes (Dynamic AI Paywall)}\\
\midrule
 Dynamic pricing/ News Distribution and Business Operations & Advanced data analytics understand users' habits, needs, and preferences to design flexible pricing structures. Similar to surge pricing used by ride-sharing services, allows news organizations to optimize business efficiency by charging prices that reflect current demand or user's specific propensity to pay. Newsrooms often look to online retail as model for how machine learning systems can automate customer journey analysis and title generation to support pricing models. & Aspirational / Reshaping the Industry: Ranked as one of top ten ways newsrooms expect AI to reshape journalism in future. While common in other sectors, remains state-of-the-art application that newsrooms currently striving to learn and implement. & \cite{beckett2019new} & \href{https://digiday.com/media/the-atlantic-introduces-dynamic-paywall-with-varying-prices-as-it-hopes-to-attract-1-million-subscribers/}{The Atlantic (Variable Pricing Smart Meter)}, \href{https://digiday.com/media/the-financial-times-ai-paywall-drove-conversions-up-290-now-its-learning-who-stays/}{Financial Times (AI-based pricing selection)}\\
\midrule
 Content Moderation/ News Distribution and Deliberation & Machine learning trained on existing bodies of moderated content builds models that classify comments as good or bad. Tools sort interactions into categories (publishable, to be rejected, potentially unsafe), allowing human moderators to focus limited time only on most problematic content. Advanced applications notify users in real-time if language likely to be flagged, encouraging self-correction. While high levels of accuracy possible (e.g., 99.7\% success rate in identifying safe comments), significant risks remain including false positives where legitimate speech wrongly censored and fundamental inability of algorithms to understand sarcasm, subtle insults, or cultural context. & Implemented / High-Priority Success: Viewed as clear-cut success for increasing newsroom efficiency and ranked among top ten ways AI expected to reshape industry. As newsrooms grapple with overwhelming volume of uncivil discourse (up to 20\% of comments may contain incivility), AI becoming essential tool to prevent total closure of commentary sections. & \cite{beckett2019new, hansen_et_al_initial_2023} & \href{https://digiday.com/media/new-york-times-moderates-12000-comments-day/}{}, \href{https://www.wired.com/2017/02/googles-troll-fighting-ai-now-belongs-world/}{The Guardian (Perspective API experiments to improve moderation), The Economist (Perspective API testing to surface toxic content faster, create safe debate space)}\\
\bottomrule
\end{longtable}

\section{Supplementary Material: Search Strategy and Screening Phases}
\label{Supplementary Material: Search Strategy and Screening Phases}
This appendix documents the complete search strategy and screening process employed in our critical narrative review.
We searched major scholarly databases, such as Scopus, ACM Digital Library, IEEE Xplore, Web of Science, JSTOR, and Google Scholar, as well as grey literature sources including newspaper articles, blog posts, technical reports, policy documents, and academic theses.
\subsection{Screening Phases}
Our search proceeded through four phases, with each phase building on findings from the previous one:
\textbf{Phase 1: FAccT $\cap$ Digital Journalism $\cap$ AI/LLMs}. Initial searches combined fairness, accountability, and transparency concepts with journalism and AI contexts. Results: 10 records retained after title/abstract screening.
\textbf{Phase 2: Discovery of Authority as Analytical Object.} Full-text review and backward reference mining revealed authority as the underlying structural condition. We pivoted search strategy toward authority, power, and control in journalism. Results: 32 records retained after full-text analysis.
\textbf{Phase 3: Authority $\cap$ Journalism.} Searches refined to editorial authority and editorial judgment as specific analytical objects. Results: 18 records retained after full-text analysis.
\textbf{Phase 4: Editorial Authority $\cap$ AI/LLMs $\cap$ Journalism.} Targeted searches intersecting editorial authority with AI/LLM adoption, supplemented by forward citation tracking from key works on algorithmic authority, platform dependencies, algocracy, and participatory methods. Results: 145 records retained after full-text analysis.
%\subsection{Inclusion and Exclusion Criteria}
%\textbf{Included:} Studies addressing at least one dimension of editorial authority relocation (decision rights, epistemic warrant, responsibility) in AI/LLM-mediated journalism; empirical studies, theoretical work, critical analyses, and practitioner accounts.
%\textbf{Excluded:} Studies treating AI as decontextualized technical artifact; non-editorial newsroom automation; general AI ethics without journalistic grounding; technical papers on model performance without authority implications.
\subsection{Inclusion and Exclusion Criteria}

The inclusion and exclusion criteria evolved alongside the screening phases as the analytical focus of the review became progressively refined. As the search moved from broad FAccT-related concerns toward editorial authority as the central analytical object, the criteria were correspondingly clarified to retain sources engaging the organization and redistribution of editorial authority in AI-mediated journalism.
We included sources addressing at least one dimension of editorial authority relocation (\emph{decision rights}, \emph{epistemic warrant}, or \emph{responsibility}) in the context of AI- or algorithmically mediated journalism. More specifically, to clarify how these criteria were operationalized during screening, we retained work examining how editorial decision-making authority (who determines publication content), epistemic warrant (how journalistic knowledge claims and verification practices are justified), or responsibility (how accountability for editorial outcomes is allocated) are configured or redistributed when AI systems, algorithms, or large language models are introduced into newsroom environments. Eligible sources therefore included empirical studies, theoretical and conceptual scholarship, critical analyses, practitioner accounts, and relevant grey literature engaging these dynamics.
This also included selected scholarship from adjacent domains, such as human–computer interaction, participatory design, platform studies, and algorithmic governance, when these works provided analytical insight into how authority, control, or responsibility are structured in sociotechnical systems that increasingly underpin journalistic production. These sources were retained as conceptual or analytical resources for interpreting authority relocation in journalism rather than as empirical studies of newsroom practice. Foundational theoretical works identified during screening were likewise retained when they directly informed the conceptualization of editorial authority used in the analysis.
Consistent with the original criteria, we excluded studies that treated AI primarily as a decontextualized technical artifact without engagement with editorial practices or authority structures in journalism. This included: (1) operational newsroom automation unrelated to editorial decision-making; (2) general discussions of AI ethics, fairness, or governance that could not support reflection on editorial practices, newsroom decision-making, or authority structures in news production; and (3) technical research on model performance or system design that framed AI and alignment solely in computational terms without considering the construction of preference data, value assumptions, governance arrangements, or their implications for editorial workflows and accountability.

\subsection{Data Synthesis}
Initial coding identified patterns of authority redistribution. These coalesced into three themes forming our organizing framework: (1) internal migrations of editorial authority; (2) external migrations to platforms, vendors, and technology companies; (3) participatory methods as counterforces. The final corpus yielded 209 records (including one methodology reference \cite{sukhera2022narrative} and 3 records added post-revisions \cite{kellogg2020algorithms, petre2015traffic, Christin2020}).

\subsection{Search Queries by Phase}
\label{tab:search-queries}
This table presents representative search queries for each phase:

\begin{longtable}
{p{0.15\linewidth}p{0.15\linewidth}p{0.60\linewidth}}
\caption{Example of records of search queries in the different screening phases}\\
\toprule
\textbf{Phase} & \textbf{Phase Description \& Rationale} & \textbf{Example Query for Database}\\
\midrule
\endhead

Phase 1: FAccT $\cap$ Digital Journalism $\cap$ AI/LLMs &
Investigated how FAccT concerns manifest in AI-mediated journalism. Search terms combined fairness, accountability, and transparency concepts with journalism and AI contexts. Results: 10 records retained after title/abstract screening. &
\textbf{Scopus:} \texttt{TITLE-ABS-KEY((("artificial intelligence" OR ai OR "machine learning" OR "large language model*" OR llm OR "generative ai") AND (journalism OR news OR newsroom*) AND (fairness OR accountability OR transparency OR bias OR ethics)))}; \textbf{ACM Digital Library:} \texttt{[[Abstract: "artificial intelligence"] OR [Abstract: "machine learning"] OR [Abstract: "generative AI"]] AND [[Abstract: journalism] OR [Abstract: news]] AND [[Abstract: fairness] OR [Abstract: accountability] OR [Abstract: transparency]]}; \textbf{IEEE Xplore:} \texttt{("Abstract":"artificial intelligence" OR "Abstract":"machine learning" OR "Abstract":"generative AI") AND ("Abstract":journalism OR "Abstract":news) AND ("Abstract":fairness OR "Abstract":accountability OR "Abstract":transparency)}; \textbf{Web of Science:} \texttt{TS=(("artificial intelligence" OR "machine learning" OR "generative AI") AND (journalism OR news OR newsroom*) AND (fairness OR accountability OR transparency OR bias))};
\textbf{JSTOR:} \texttt{(ab:("artificial intelligence" OR "machine learning" OR "generative AI") AND ab:(journalism OR news) AND ab:(fairness OR accountability OR transparency))};
\textbf{Google Scholar:} \texttt{"artificial intelligence" OR "machine learning" OR "generative AI" journalism news fairness OR accountability OR transparency}\\
Phase 2: Discovery of Authority as Analytical Object &
Database search, full-text review, and backward reference mining revealed FAccT concerns repeatedly manifested through struggles over decision rights, epistemic warrant, and responsibility. Authority emerged as the underlying structural condition determining fairness (whose perspectives matter), accountability (who is responsible), and transparency (who can audit). We pivoted search strategy to focus on authority, power, and control in journalism.  Results: 32 retained after full-text analysis. &
\textbf{Scopus:} \texttt{TITLE-ABS-KEY(((authority OR power OR control OR "decision rights") AND (journalism OR news OR "news production" OR newsroom*)))}; \textbf{ACM Digital Library:} \texttt{[[All: authority] OR [All: power] OR [All: control]] AND [[All: journalism] OR [All: news] OR [All: newsroom]]}; \textbf{IEEE Xplore:} \texttt{("All Metadata":authority OR "All Metadata":power OR "All Metadata":control OR "All Metadata":"decision rights") AND ("All Metadata":journalism OR "All Metadata":newsroom)};
\textbf{Web of Science:} \texttt{TS=((authority OR power OR control OR "decision making" OR "decision rights") AND (journalism OR news OR newsroom* OR "news production"))};
\textbf{JSTOR:} \texttt{((ti:(authority OR power OR control) OR ab:(authority OR power OR control)) AND (ti:(journalism OR newsroom OR news) OR ab:(journalism OR newsroom OR news)))}; \textbf{Google Scholar:} \texttt{authority OR power OR control journalism newsroom "news production"}\\
Phase 3: Authority $\cap$ Journalism &
Searched for authority, power, and control in journalistic contexts. Identified editorial authority and editorial judgment as the specific forms of institutionally-recognized authority at stake—the capacity to make binding decisions about publication content, distinct from general organizational authority. Results: 18 records retained after after full-text analysis. &
\textbf{Scopus:} \texttt{TITLE-ABS-KEY((("editorial authority" OR "journalistic authority" OR gatekeeping OR "editorial control") AND (journalism OR news OR newsroom*)))}; \textbf{ACM Digital Library:} \texttt{[[All: "editorial authority"] OR [All: "journalistic authority"] OR [All: gatekeeping]] AND [[All: journalism] OR [All: newsroom]]}; \textbf{IEEE Xplore:} \texttt{("All Metadata":"editorial authority" OR "All Metadata":"journalistic authority" OR "All Metadata":gatekeeping) AND ("All Metadata":journalism OR "All Metadata":newsroom)}; \textbf{Web of Science:} \texttt{TS=(("editorial authority" OR "journalistic authority" OR gatekeeping OR "editorial control") AND (journalism OR news OR newsroom*))}; \textbf{JSTOR:} \texttt{(ti:("editorial authority" OR "journalistic authority" OR gatekeeping) OR ab:("editorial authority" OR "journalistic authority" OR gatekeeping)) AND (ti:(journalism OR newsroom) OR ab:(journalism OR newsroom))}; \textbf{Google Scholar:} \texttt{"editorial authority" OR "journalistic authority" OR gatekeeping journalism newsroom}\\
Phase 4: Editorial Authority $\cap$ AI/LLMs $\cap$ Journalism &
With editorial authority established as our analytical object, we conducted targeted searches intersecting these concepts with AI/LLM adoption. Supplemented database searches with forward citation tracking from key works on algorithmic authority, platform dependencies, algocracy, and participatory methods. Results: 145 records retained after full-text analysis. &
\textbf{Scopus:} \texttt{TITLE-ABS-KEY((("editorial authority" OR "journalistic authority" OR "editorial judgment") AND ("artificial intelligence" OR ai OR "machine learning" OR llm OR algorithm*) AND (journalism OR news OR newsroom*)))}; \textbf{ACM Digital Library:} \texttt{[[All: "editorial authority"] OR [All: "journalistic authority"]] AND [[All: "artificial intelligence"] OR [All: "machine learning"] OR [All: algorithm]] AND [[All: journalism] OR [All: newsroom]]}; \textbf{IEEE Xplore:} \texttt{("All Metadata":"editorial authority" OR "All Metadata":"editorial judgment") AND ("All Metadata":"artificial intelligence" OR "All Metadata":algorithm*) AND ("All Metadata":journalism OR "All Metadata":newsroom)}; \textbf{Web of Science:} \texttt{TS=(("editorial authority" OR "journalistic authority" OR "editorial judgment") AND ("artificial intelligence" OR "machine learning" OR llm OR algorithm*) AND (journalism OR news OR newsroom*))}; \textbf{JSTOR:} \texttt{((ti:("editorial authority" OR "journalistic authority") OR ab:("editorial authority" OR "journalistic authority")) AND (ti:("artificial intelligence" OR algorithm*) OR ab:("artificial intelligence" OR algorithm*)) AND (ti:(journalism OR newsroom) OR ab:(journalism OR newsroom)))}; \textbf{Google Scholar:} \texttt{"editorial authority" OR "journalistic authority" "artificial intelligence" OR algorithm journalism newsroom};
\textbf{Scopus (supplementary):} \texttt{TITLE-ABS-KEY((("algorithmic authority" OR "platform dependenc*" OR "editorial control") AND ("ai adoption" OR "ai integration" OR automation) AND (journalism OR newsroom*)))};\textbf{Web of Science (supplementary):} \texttt{TS=((authority OR power OR control OR "decision making") AND ("generative ai" OR llm OR "language model*") AND (journalism OR "news production" OR newsroom*))};\textbf{Google Scholar (supplementary):} \texttt{"platform dependency" OR "algorithmic authority" "AI adoption" journalism newsroom}\\
\bottomrule
\end{longtable}

\clearpage

\subsection{Distribution of retained records across screening phases}
\label{screenin-phases-references}
This table presents the distribution of retained sources across the different screening phases: 
\begin{longtable}{p{0.12\linewidth}p{0.50\linewidth}p{0.28\linewidth}}
\caption{Distribution of retained sources across screening phases with reference entries}
\label{tab:screening-phases-references}\\
\toprule
\textbf{Phase} & \textbf{Description} & \textbf{References} \\
\midrule
\endhead

Phase 1 (n=10) & FAccT $\cap$ Digital Journalism $\cap$ AI/LLMs: Algorithmic accountability, transparency limitations, fairness in AI systems & \cite{ananny2018seeing, bender_dangers_2021, cooper_accountability_2022, diakopoulos2015algorithmic, gorwa2020algorithmic, knowles_sanction_2021, neumann2022justice, 10.1145/3715275.3732052, raji2020closing, selbst2019fairness} \\
\addlinespace

Phase 2 (n=32) & Authority/Power $\cap$ Journalism: Power structures, organizational control, datafication, authorship theory & \cite{ananny_2018_freedom, anderson2013towards, barthes_death_1997, coddington_clarifying_2015, cushion_data_2017, deuze2008changing, dosi_hierarchies_2021, ausserhofer_et_al_datafication_2020, farrell_moral_2023, foucault1977what, foucault1980power, foucault_truth_2009, garcia_aviles_journalists_2004, hayles_2002_print, hayles_how_2010, hayles_cognitive_2016, hilligoss_developing_2008, humprecht_ownership_2019, jungherr2021digital, Kitchin02012017, kosterich2020managing, lewis2015actors, marvin_when_1988, oldham_navigating_2025, pavlik_impact_2000, punday_2025_digital, schuster_power_2024, seaver_algorithms_2017, skrastins_how_2019, stalph_classifying_2018, tabary2016data, turow2013daily} \\
\addlinespace

Phase 3 (n=18) & Editorial Authority $\cap$ Journalism: Gatekeeping, journalistic autonomy, editorial judgment, professional boundaries & \cite{breed_social_1955, carlson_journalistic_2017, carlson2020journalistic, carlson_boundaries_2015, gans_deciding_2004, hamada2019editorial, hamada_determinants_2022, hassan_journalism_2024, jane_b_singer_contested_2007, karppinen_what_2016, reich_and_hanitzsch_full_2013, schudson_virtues_2005, shoemaker2009gatekeeping, smeenk_journalist_2023, tuchman1972objectivity, tuchman_making_1978, wallace2017modelling, zilberstein_models_2024} \\
\addlinespace

Phase 4 (n=145) & Editorial Authority $\cap$ AI/LLMs + Participatory Methods + Platform Dependencies: AI adoption in newsrooms, human-AI interaction, participatory design, platform power, algorithmic authority & 

\cite{ahmad_openais_2025, Amigo20042025, ajmani_secondary_2025, akbulut2024all, albizu-rivas_artificial_2024, amershi2019guidelines, becker2023new, amponsah2024navigating, aneesh2006virtual, annapureddy_generative_2025, arnstein_ladder_1969, bai_constitutional_2022, balayn_unpacking_2025, bannon_reimagining_2018, becker_policies_2025, beckett2019new, beckett2023generating, bell_platform_2017, bennett2023does, birhane_power_2022, bodo_selling_2019, bomba_agency_2025, bommasani_opportunities_2021, bratteteig_disentangling_2012, carlson_robotic_2015, carlson2018automating, carnat2024human, chu_user_2026, chua2022platform, cooke_participation_2001, cools_towards_2023, cools_uses_2024, cools_news_2024, cooper_systematic_2022, cooper_fitting_2024, corbett_power_2023, delgado_stakeholder_2021, delgado_participatory_2023, diakopoulos2014algorithmic, diakopoulos_automating_2019, diakopoulos2020computational, diakopoulos_generative_2024, dierickx_pdf_2021, dodds2024impact, dodds_controlled_2025, dorr2017ethical, drunen_safeguarding_2023, elish_moral_2019, eslami2015always, hansen_et_al_initial_2023, lewis_et_al_pdf_2019, feffer_preference_2023, fischer_understanding_2011, fischer2025efficiency, friedman_value_2003, fritz_deference_2025, gabriel_artificial_2020, gillespie2014relevance, gordon_jury_2022, groves_going_2023, gutierrez_lopez_question_2023, hopkins_recourse_2025, hoque_towards_2024, huang_collective_2024, inie_designing_2023, jakesch2023co, jones_ai_2022, jones2023generative, kapania2022because, khamassi_strong_2024, kirk_prism_2024, klingefjord_what_2024, komatsu2020ai, konya_democratic_2023, kunert_form_2019, lai_towards_2021, lee_webuildai_2019, lermann2024effects, levy2021assessing, lewis_generative_2025, liel2025turning, lischka2023editorial, liu2016reuters, longoni_news_2021, luger_like_2016, malcorps_news_2019, matthias2004resp, mazeika_utility_2025, moran_robots_2022, moller_reinforce_2024, nass_computers_1994, nayebare2023openai, nekoto_participatory_2020, nielsen2022power, nishal_envisioning_2024, norman_how_1994, medill2024impact, oh2025harmonizing, outi_lundahl_algorithmic_2022, paik_journalism_2025, palacin_design_2020, papaevangelou_funding_2024, parasuraman_complacency_2010, parasuraman_humans_1997, parasuraman2000model, peter2025benefits, petridis2023anglek, poell2023spaces, pushkarna_data_2022, rader2015understanding, rafailov_direct_2024, rakova_where_2021, reeves1996media, romeo2025exploring, salikutluk2024evaluation, santoni2018meaningful, schapals_assistance_2020, schoeffer2024explanations, schoeffer2025ai, shin_automating_2025, shneiderman2020human, simon_uneasy_2022, simon_artificial_2024, simon_escape_2024, sloane_participation_2022, sonni_digital_2024, sorensen_value_2024, spangher-etal-2024-llms, sundar2019machine, suresh_participation_2024, thasler2025journalistic, thurman_algorithms_2019, tsamados_human_2025, tseng_ownership_2025, european_broadcasting_union_leading_2025, vaccaro2021contestability, van_dalen_revisiting_2024, van_drunen_editorial_2021, wang_reelframer_2024, john_wihbey_ai_2024, wu_journalists_2024, wu_honor_2023, xiao_it_2025, zhou2023synthetic}
\\
\addlinespace

Methodology (n=1) & Narrative review methods & \cite{sukhera2022narrative} \\
\addlinespace
Post revision additions (n=3) & Sources on algorithmic control and newsroom metrics & \cite{kellogg2020algorithms, Christin2020, petre2015traffic} \\
\addlinespace
\textbf{Total} & \textbf{Final Corpus} & \textbf{n=209} \\
\bottomrule
\end{longtable}

\subsection{Distribution of records across years and thematic category} \label{records distribution}

\begin{figure*}
    \centering
    \includegraphics[width=1\linewidth]{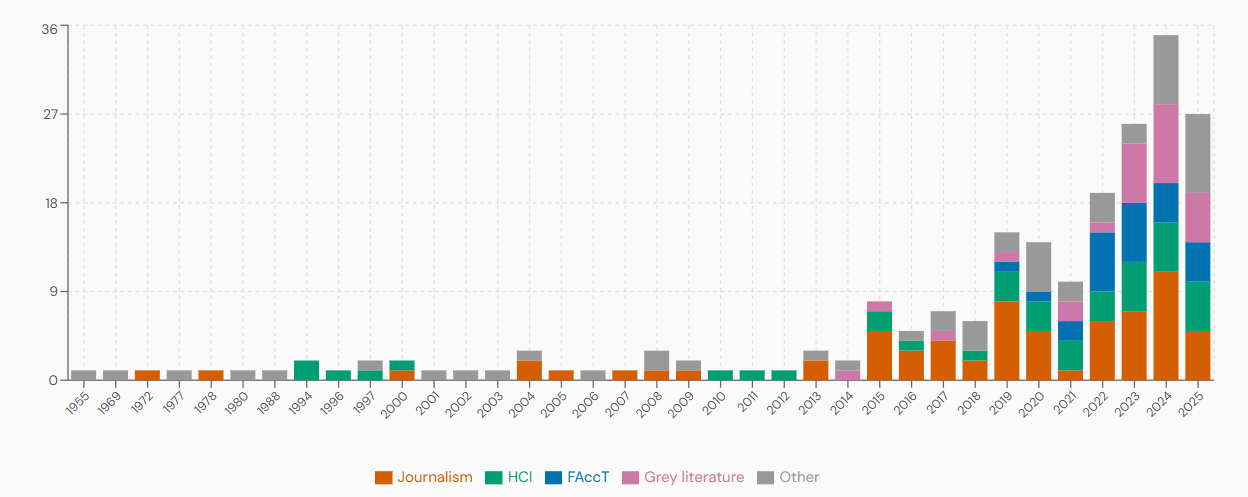}
    \caption{Distribution of the 209 references by publication year and thematic category. The corpus spans seven decades (1955–2025). The temporal distribution is heavily concentrated in the recent period: 147 references (70.3\%) were published from 2019 onward, reflecting the rapid growth of scholarship at the intersection of AI and journalism, HCI, and algorithmic accountability. The years 2024 (n = 35) and 2025 (n = 28) represent the two largest single-year cohorts. A further 49 references (23.4\%) were published in the 2010s, while 28 (13.4\%) predate 2010 and include foundational works in media theory, philosophy, sociology, and early HCI. References were first divided into peer-reviewed or book-based sources (183; 87.6\%) and grey literature (26; 12.4\%). Grey literature encompasses non-peer-reviewed or institutionally published material such as technical reports, white papers, arXiv preprints, working papers, theses, and institutional reports, and was assigned regardless of topical fit, so that, for example, an arXiv manuscript on journalism or AI governance was classified as grey literature rather than under a thematic heading. Within the peer-reviewed corpus, each reference was assigned to a single thematic category using a priority rule based on venue, title, and object of analysis. Journalism (n = 68; 32.5\%) is the largest category and includes work centered on newsrooms, editorial workflows, journalistic authority and autonomy, computational journalism, automation in news, and media–platform dynamics, typically published in venues such as Digital Journalism, Journalism Practice, and Journalism Studies. HCI (n = 39; 18.7\%) groups work on human–AI interaction, user studies, participatory design, trust, reliance, explainability in use, and collaborative work practices, primarily from CHI, CSCW, and IJHCS-type venues. FAccT (n = 24; 11.5\%) covers fairness, accountability, transparency, participatory AI, governance, and stakeholder harms, especially from FAccT, EAAMO, and AIES-type venues. Other (n = 52; 24.9\%) serves as a residual category for foundational and adjacent theory, including STS, philosophy, sociology, organizational studies, general media theory, authorship theory, and general AI ethics, that informs the review but does not fit cleanly into the preceding categories. Ambiguous cases were resolved by prioritising venue and core contribution: a CHI paper touching on AI governance was classified as HCI if its primary contribution concerned interaction or design, while a FAccT paper with HCI elements was classified as FAccT if its central framing was accountability or socio-technical justice.}
    \label{fig:distribution}
\end{figure*}

\clearpage
\section{Supplementary Material: Representative Discussion Quotes by Theme}
\label{Supplementary Material: Representative Discussion Quotes by Theme}
The table in this appendix presents representative quotations used to substantiate the core claims developed in the Discussion and Conclusion section (\ref{discussion}), grouped according to the three analytical themes of the review. The aim is to make the interpretive basis of the discussion more transparent by presenting examples of the textual evidence underpinning each claim. The excerpts reported here were selected because they explicitly refer to the journalistic domain and they provide relevant conceptual or empirical support from the cited sources, while indicating, where necessary, whether a quotation reflects an original authorial formulation or a synthetic statement grounded in prior scholarship.

\begin{longtable}{p{0.15\linewidth}p{0.24\linewidth}p{0.10\linewidth}p{0.40\linewidth}}
\caption{Supporting references and representative quotes for the core discussion claims, subdivided across the three themes}\\
\toprule
\textbf{Theme} & \textbf{Claim} & \textbf{Source} & \textbf{Representative Quote(s)}\\
\midrule
\endhead

\textbf{Theme 1: Internal Migrations of Editorial Authority}
& Research on the digital transformation of newsrooms shows how editorial authority increasingly emerges from socio-technical assemblages in which computational systems participate in shaping editorial routines and decision environments.
& \texttt{\cite{diakopoulos_automating_2019}}
& ``Actor network theory (ANT) considers behavior as emerging from an assemblage of humans (actors), objects and technologies (actants), and their relationships.'' \newline
``For computational journalism it is essential to recognize the range of the actors and actants---both human and nonhuman, and inside or outside the newsroom---and examine the associations they engage in as news information is produced.'' \newline
``We must ask not only `who' does journalism, but also `what' does journalism, and that `what' includes technical artifacts and algorithms.'' \\

& & \texttt{\cite{thasler2025journalistic}}
& ``... when journalists share their editorial authority with opaque algorithmic systems, the power relations in the production of news and public knowledge change.'' \newline
``The diffusion of power between human and technological agents can manifest in how their contributions determine the collaboration’s outcome.'' \newline
(\textit{Note: The underlying argument of the quotes is grounded by the authors in prior scholarship.}) \\

& & \texttt{\cite{seaver_algorithms_2017}}
& ``...I saw [the term]`algorithms' mutate, at times encompassing teams of people and their decisions, and at other times referring to emergent, surprisingly mysterious properties of the codebase.'' \newline
``In practice, [the term] `algorithm' had a vague, `non-technical' meaning, indicating various properties of a broader `algorithmic system', even in nominally `technical' settings.'' \\

\midrule

& Work on human–AI interaction further demonstrates how generative systems acquire perceived epistemic legitimacy through fluency, stylistic competence, and social cues aligned with professional journalistic practices.
& \texttt{\cite{peter2025benefits}}
& ``A growing body of research suggests that the recent generation of large language model (LLMs) excel, and in many cases outpace humans, at writing persuasively and empathetically, and at mimicking human-like conversation believably and effectively.'' \newline
``...the latest generation LLMs have encoded language at such nuanced levels as to mimic highly convincingly a range of communicative abilities...'' \newline
``LLMs excel at role-play, capable of impersonating a wide-range of roles and personas...'' \newline
(\textit{Note: The underlying argument of the quotes is grounded by the authors in prior scholarship.}) \\

& & \texttt{\cite{reeves1996media}}
& ``Computers, in the way that they communicate, instruct, and take turns interacting, are close enough to human that they encourage social responses.'' \newline
``When it comes to being social, people are built to make the conservative error: When in doubt, treat it as human.'' \\

& & \texttt{\cite{carnat2024human}}
& ``...due to their highly persuasive and human-like answers... users may be prone to overrelying on the chatbot’s generated output in their decision-making...'' \newline
``Research suggests that the more human-like a system appears, the higher the likelihood of the user attributing human traits and capabilities to it...'' \newline
(\textit{Note: The underlying argument of the quotes is grounded by the authors in prior scholarship.}) \\

& & \texttt{\cite{zhou2023synthetic}}
& ``...AI-misinfo cleverly mimicked the attributes of existing information assessment guidelines, thus giving false impressions of their veracity.'' \newline
``By pinpointing researcher names, institutions... publication venues... and expertise/reputation... AI-misinfo tended to appear credible, and potentially, more persuasive to readers.'' \\

\midrule

& Alignment practices and institutional voice mimicry further allow LLM outputs to reproduce journalistic norms and values, creating perceptions of institutional alignment that exceed the models’ normative grounding.
& \texttt{\cite{khamassi_strong_2024}}
& ``...we have proposed a novel distinction between weak and strong alignment: the former corresponding to statistically aligned behavior without understanding what human values are, mean or imply...'' \newline
``We further argue that strong alignment itself requires... an understanding of what human values are...'' \\

& & \texttt{\cite{mazeika_utility_2025}}
& ``...reward models trained by human feedback do represent a form of `utility' for guiding generated text, they should not be conflated with an LLM’s own internal values...'' \newline
``In contrast to alignment methods that modify surface behaviors through a noisy human reward proxy... utility control aims to directly reshape the underlying preference structures...'' \\

& & \texttt{\cite{jakesch2023co}}
& ``The results also show how simple it is to make models highly opinionated using accessible methods like prompt engineering.''\\

& & \texttt{\cite{sundar2019machine}}
& ``...the most immediately accessible ideas about machines are the ones that are well-entrenched in our mental models, which tend to be the positive stereotypes about machine infallibility and neutrality.'' \\

\midrule

& Scholarship on oversight in AI-mediated workflows shows how these dynamics generate responsibility gaps in which journalists retain formal accountability while exercising reduced control over the epistemic processes producing content.
& \texttt{\cite{matthias2004resp}}
& ``Now it can be shown that there is an increasing class of machine actions... because nobody has enough control over the machine’s actions to be able to assume the responsibility for them.'' \newline
``These cases constitute what we will call the responsibility gap.'' \\

& & \texttt{\cite{elish_moral_2019}}
& ``I articulate the concept of a moral crumple zone to describe how responsibility for an action may be misattributed to a human actor who had limited control over the behavior of an automated or autonomous system.'' \newline
``The moral crumple zone protects the integrity of the technological system, at the expense of the nearest human operator.'' \newline
``...[there are] incongruities between control and responsibility...'' \\

& & \texttt{\cite{schoeffer2025ai}}
& ``...AI systems often serve as decision support tools that aid human experts, who retain discretionary power to override AI recommendations and ultimately bear responsibility for making final decisions.'' \newline
``In practice, however, we often see that humans cannot assess the correctness of AI recommendations...'' \\

& & \texttt{\cite{dodds_controlled_2025}}
& ``As AI-generated content becomes more sophisticated, it challenges the foundations of journalistic expertise...'' \newline
``...news professionals must reposition themselves not as mere content creators but as ethical supervisors of AI-generated content...'' \\

\midrule

\textbf{Theme 2: External Migrations of Editorial Authority}
& Research on the political economy and organizational structure of digital media shows that editorial authority has long been shaped by dependencies on funding structures, infrastructures, ownership arrangements, and distribution systems, alongside hierarchical newsroom governance and contested editorial independence.
& \texttt{\cite{breed_social_1955}}
& ``The publisher ordinarily owns the paper and from a purely business standpoint has the right to expect obedience of his employees. He has the power to fire or demote for transgressions.'' \newline
``Basically, the learning of policy is a process by which the recruit discovers and internalizes the rights and obligations of his status and its norms and values. He learns to anticipate what is expected of him so as to win rewards and avoid punishments.'' \\

& & \texttt{\cite{humprecht_ownership_2019}}
& ``Owning a news outlet can give owners the power to use the news produced for their own purposes. Some owners might try to intervene in news production in order to promote their own opinions and viewpoints. Alternatively, they aim for profit maximization and thus promote the production of easy-to-sell content.'' \\

& & \texttt{\cite{drunen_safeguarding_2023}}
& ``Definitions of editorial independence in both legal and journalism studies literature broadly focus on the ability to determine how news is produced, published, and distributed to the public free from outside influences... complete editorial independence is neither possible nor desirable... The media has always had to rely on others to for instance fund journalism, access information, and implement technology.'' \newline
``The technical expertise and resources needed to automate editorial decision-making can moreover create new dependencies with regard to other parties, such as engineers, product owners, third-party technology companies, and data brokers.'' \newline
(\textit{Note: The underlying argument of the quotes is grounded by the authors in prior scholarship.}) \\

& & \texttt{\cite{simon_uneasy_2022}}
& ``Platform companies have already wrought control and autonomy from the news as they have taken over the dominant revenue model (advertising) and as news organisations have adopted and submitted to the logics of their platforms, their distribution systems, and ways of measuring and targeting audiences.'' \newline
``While the power exercised by platform companies over the news has to date mainly flown from their control over the channels of communication and distribution, AI allows them to extend this control as the technology increasingly permeates all stages of the gatekeeping process from news gathering, to production, to distribution.'' \\

\midrule

& Studies of platformization, algorithmic governance, and infrastructure provision show that AI adoption intensifies these dynamics by shifting decision-making power toward platforms, vendors, and technology companies that increasingly provide the tools, infrastructures, and distribution channels used in news production.
& \texttt{\cite{papaevangelou_funding_2024}}
& ``Both platforms’ endeavors to fund journalism, news organizations and other relevant stakeholders are part and parcel of their strategy to become the de-facto infrastructural backbone of the news industry.'' \\

& & \texttt{\cite{simon_escape_2024}}
& ``...the introduction of AI reshapes the dependency of publishers on platform companies by exacerbating existing dependencies on the distribution side and introducing new dependencies on the production side, providing platform companies with greater infrastructural control over news organisations.'' \newline
``On the macro level, increasing platform control through AI likely further strengthens an already asymmetric power relationship... with a handful of platform companies holding outsized control over the direction of the technology and the conditions for its use.'' \newline
``While publishers can innovate by using platforms’ AI, they can only do so within the constraints set by the same given that platform firms possess artefactual and contractual control over their technology and infrastructure.'' \\

& & \texttt{\cite{poell2023spaces}}
& ``With the widespread uptake of digital platforms and apps, news organizations around the globe have progressively aligned their business models, distribution infrastructures, and production practices with platforms.'' \newline
``...platform companies predominantly set the terms and conditions under which news organizations can produce, distribute, and monetize content... In short, news organizations are much more dependent on platforms than the converse.'' \newline
``...news content distribution, marketing, and monetization are increasingly integrated with platform infrastructures.'' \newline
(\textit{Note: The underlying argument of the quotes is grounded by the authors in prior scholarship.})\\

& & \texttt{\cite{bell_platform_2017}}
& ``Technology platforms have become publishers in a short space of time, leaving news organizations confused about their own future.'' \newline
``Technology companies including Apple, Google, Snapchat, Twitter, and, above all, Facebook have taken on most of the functions of news organizations, becoming key players in the news ecosystem... The distribution and presentation of information, the monetization of publishing, and the relationship with the audience are all dominated by a handful of platforms.'' \\

\midrule

& Research on infrastructural capture and platform dependence further highlights how technical capabilities, datasets, and computational resources become concentrated among a small number of actors while external optimization logics are embedded into editorial processes.
& \texttt{\cite{simon_escape_2024}}
& ``On the macro level, increasing platform control through AI likely further strengthens an already asymmetric power relationship... with a handful of platform companies holding outsized control over the direction of the technology and the conditions for its use.'' \newline
``Publishers rely on platform companies for AI development... due to high costs and resource constraints, including limited data, expertise, labor, and computing power.'' \newline
``AI increases platforms’ capacity to embed their own logics---of greater quantification and commercialisation...---on the journalistic field.'' \newline
(\textit{Note: The underlying argument of the third quote is grounded by the authors in prior scholarship.}) \\

& & \texttt{\cite{simon_artificial_2024}}
& ``Given the nature of the technology---which has a high barrier to entry due to the vast data and computational power requirements---platform companies are likely to dominate this space for the foreseeable future.'' \newline
``AI doesn’t just continue this dynamic: it intensifies it. It shifts newswork even further toward the technical and the logics of platform companies: prioritizing greater rationalization and calculability... and efficiencies and productivity...'' \newline
``...the (im)balance of power will tip ever more toward the technology companies that provide access to them---at the expense of the news organizations whose journalism has been used to train and improve them.'' \\

& & \texttt{\cite{hansen_et_al_initial_2023}}
& ``...it places immense power in the hands of very few companies... these are the five `tech giants' Google, Facebook (now Meta), Microsoft, Apple, and Amazon who predominately provide both data and other technical infrastructures for... the media sector.'' \newline
``...the continued reliance on external data collections systems or readymade solutions of recommender system from external parties... raise critical questions, because media organisations are becoming reliant on existing (often commercially oriented) measures and systems.'' \newline
\\

& & \texttt{\cite{petre2015traffic}}
& ``In both popular and academic discourse, there is a widespread tendency to think of metrics as instruments of rationalization---as tools that systematically and scientifically measure the performance of news content...'' \newline
``Writers and editors had a tendency to go with what works, or had proven to work in the past... In the event that a post outside the realm of surefire traffic-getters became a surprise hit, a site would try to follow up on it or replicate it.'' \\

\midrule

& These (external) dependencies also reshape internal newsroom roles and decision structures, redistributing authority toward technical specialists, managers, and algorithmic systems while narrowing journalists’ autonomy and fragmenting accountability within digitized workflows. 
& \texttt{\cite{garcia_aviles_journalists_2004}}
& ``Media managers take charge of the traffic of information inside the newsroom... Some responsibility for content decisions, although nominally resting with the programme editor, has been transferred in practice to the media manager.'' \newline
``Some new positions have been introduced in the traffic control room... some of the new technical jobs involve little creativity and a great deal of responsibility.'' \\

& & \texttt{\cite{van_drunen_editorial_2021}}
& ``Editorial control, which has long been a necessarily human activity, is now ceded in whole or in part to algorithms (Diakopoulos, 2019, p. 28)... This creates new tensions regarding the ways in which editors and journalists can continue to define and safeguard editorial values.''\newline
(\textit{Note: The underlying argument of the quotes is grounded by the authors in prior scholarship.}) \\

& & \texttt{\cite{cools_news_2024}}
& ``Scholars have pointed out how automation is taking over tasks traditionally executed by journalists, such as analyzing data to identify patterns and deciding which information to highlight or neglect... This integration of automated editorial processes has led to increased efficiency but also decreased the role of human decision-making in news production.'' \newline
``These changes in the news industry have repercussions for journalistic work, impacting news workers’ authority and autonomy and may compromise journalistic norms...'' \newline
(\textit{Note: The underlying argument of the quotes is grounded by the authors in prior scholarship.}) \\

& & \texttt{\cite{dodds2024impact}}
& ``...internally, there is a prevalent attitude that assigns responsibility to other departments or individuals within the organizations... This delineation of accountability, or rather a displacement thereof, signifies an issue of responsibility fragmentation within news organizations...'' \newline
``...our interviews revealed that journalists often defer responsibility for the ethical and responsible implementation of AI in the newsroom to the tech or the legal team.'' \\

\midrule

\textbf{Theme 3: Editorial Authority Retention Through Participatory Methods in AI-Mediated Journalism} & Participatory design and responsible AI research suggests that involving affected stakeholders in the development, alignment and implementation of AI systems in professional contexts such as journalism can reshape authority relations by redistributing decision rights and epistemic control, while different degrees of involvement create varying capacities for actors to influence model behavior, data selection, and evaluation criteria.
& \texttt{\cite{arnstein_ladder_1969}}
& ``Citizen participation is a categorical term for citizen power. It is the redistribution of power that enables the have-not citizens... to be deliberately included in the future.'' \newline
``There is a critical difference between going through the empty ritual of participation and having the real power needed to affect the outcome of the process.'' \newline
``At the topmost rungs... have-not citizens obtain the majority of decision-making seats, or full managerial power.'' \\

& & \texttt{\cite{delgado_participatory_2023}}
& ``The extent to which stakeholders are able to determine which design decisions they will inform or even make is a key differentiator across the various participatory schools of thought we surveyed...'' \newline
``These four modes of participation form a spectrum... The most substantive form of participation occurs when stakeholders own the design process, playing a central role in shaping the procedures for deliberation as well as deciding on the outcomes...'' \newline
``Only 10\% (8 of 80) of the projects allowed stakeholders to provide input into the type of model and features used, the design of objective or loss functions, or... the decision thresholds.'' \\

& & \texttt{\cite{suresh_participation_2024}}
& ``If meaningful participation requires the distribution of some decision-making power, this shift is not easily managed by firms...'' \newline
``In a classic ML setup... participatory ML approaches might look like community control over problem formulation, data collection and storage, or the design of evaluations that best reflect real-world use.'' \newline
``The subfloor layer encompasses technical infrastructure, norms and governance for a grounded domain... fine-tuned models, curated datasets, auditing process, mechanisms for recourse, standards-setting procedures, and shared governance structures...'' \\

& & \texttt{\cite{ajmani_secondary_2025}}
& ``Meaningful participation is the step-by-step process of building (1) informed, (2) consensual, and (3) agentic relationships with AI...'' \newline
``Agency is about giving participants the capacity to shape the outcomes of their participation, such as resulting datasets, tools, and algorithmic recommendations.'' \newline
``...using the term `participation' without creating space for ownership creates a problematic class of AI participants who do not have any say in how their participation affects the system.'' \\

& & \texttt{\cite{tseng_ownership_2025}}
& ``...there is a critical opportunity to instead pursue a participatory approach to the development and governance of LLMs and AI, owned and led by journalists.'' \newline
``...participatory approaches toward LLMs provide three core benefits... (i) the ability to conduct holistic and highly contextual evaluations... (ii) improved coverage of diverse perspectives via responsible data sharing; and (iii) education and capacity-building.'' \\

\midrule

& Such (participatory) approaches can help retain editorial authority in newsrooms by making the epistemic infrastructure of AI systems more transparent, contestable, and institutionally grounded.
& \texttt{\cite{suresh_participation_2024}}
& ``The subfloor layer encompasses technical infrastructure, norms and governance for a grounded domain... fine-tuned models, curated datasets, auditing process, mechanisms for recourse, standards-setting procedures, and shared governance structures.'' \newline
``It breaks the participatory ceiling through being grounded in context and primarily owned by entities such nonprofits, local governments, community advocacy organizations, or unions.'' \newline
``...surface layer interventions could offer a space for individual data refusal; and subfloor layer interventions could offer a space for collective contestation around data usage.'' \\

& & \texttt{\cite{hoque_towards_2024}}
& ``Presenting information alongside explanations of how and why news stories were developed can significantly enhance the trust readers place in news outlets.'' \newline
``Successfully answering these questions about the news production process... would help increase trust and accountability.'' \newline
``...journalists prefer a human-in-the-loop process rather than delegating all questions to a chatbot...'' \\

& & \texttt{\cite{cooper_fitting_2024}}
& ``Elevating such expertise requires attending to frustration, refusal, and dissent concerning ML systems as viable forms of agency and key sources of feedback.'' \newline
``...participating with respect to deployed systems `requires a certain kind of literacy... so that you can actually navigate your data representation'.'' \newline
``...this requires the development of methodologies that enable people interacting with ML systems to express their interests in ways that can be input into ML development and related policy processes.'' \\

& & \texttt{\cite{corbett_power_2023}}
& ``AI has risks creating a state of algrocacy, `when authority becomes embedded in the technology itself'.'' \newline
``There has been a push to incorporate more participatory methods in AI developments processes in order to create more transparent, fair, and equitable systems.'' \newline
``Decisions made at this stage embed normative commitments, beliefs, and values... making it one of the most opportune... entry point for public participation.'' \\

\midrule

& Participation remains conditional: structural power asymmetries, limited resources, and risks of pseudo-participation can leave underlying authority configurations largely unchanged or even legitimized.
& \texttt{\cite{sloane_participation_2022}}
& ``...we want to caution against... `participation washing'... where `participation'... is based on manufactured consent and... global power.'' \newline
``...stakeholder engagement protocols often legitimize injustices... and systematically devalue user needs as part of profit and scale-oriented design practices...'' \newline
``Using the language of design justice without actually engaging in actual design justice processes and practices can only lead to corporate co-optation.'' \\

& & \texttt{\cite{palacin_design_2020}}
& ``...a lack of agency in decision making and agenda-setting is a growing phenomenon in the design of digital public services. We call this pseudo-participation...'' \newline
``A process can be pre-set by an external agenda and not offer any meaningful power to the people, thus leading to participation without agency.'' \newline
``...these artifact’s affordances have been pre-set by an agenda and do not offer any meaningful power to the people.'' \\

& & \texttt{\cite{birhane_power_2022}}
& ``The risk that participation could simply mask uneven power relations, making it easier to perpetuate a dynamic that is fundamentally extractive remains a major concern.'' \newline
``The heavy influence of industry stakeholders calling for greater participation without resolving these tensions has led to concerns of `participation-washing'...'' \\

& & \texttt{\cite{rakova_where_2021}}
& ``Most commonly, practitioners have to grapple with lack of accountability, ill-informed performance trade-offs and misalignment of incentives within decision-making structures...'' \newline
``...responsible AI work is often not compensated... or is perceived as ambiguous or too complicated for the organization’s current level of resources.'' \newline
``...overly rigid organizational incentives may demotivate addressing them, compounded by organizational inertia...'' \\

\bottomrule
\end{longtable}

\clearpage

\end{document}